\documentclass[12pt]{iopart}

\usepackage{iopams}  
\usepackage[utf8]{inputenc}
\usepackage{xcolor}
\usepackage[]{siunitx}

\expandafter\let\csname equation*\endcsname\relax

\expandafter\let\csname endequation*\endcsname\relax

\usepackage{amsmath}
\usepackage{pxfonts}
\usepackage{graphicx}

\usepackage{bm}
\usepackage{subcaption}
\usepackage{breqn}
\usepackage{comment}

\newcommand{\bea}{\begin{eqnarray}}
\newcommand{\eea}{\end{eqnarray}}
\newcommand*{\defeq}{\mathrel{\vcenter{\baselineskip0.5ex \lineskiplimit0pt
                     \hbox{\scriptsize.}\hbox{\scriptsize.}}}%
                     =}

\begin{document}
\title{Uncovering the non-equilibrium stationary properties in sparse Boolean networks}
\author{Giuseppe Torrisi, Reimer  K\"uhn, Alessia Annibale}
\begin{abstract}
Dynamic processes of interacting units on a network  are out of equilibrium in general.
In the case of a directed tree, the dynamic cavity method provides an efficient tool that characterises the dynamic trajectory of the process for the linear threshold model. However, because of the  computational complexity of the method, the analysis has  been limited to systems where the largest number of neighbours is small. We devise an efficient implementation of the dynamic cavity method which substantially reduces the computational complexity of the method for systems with discrete couplings. Our approach opens up the possibility to investigate the dynamic properties of networks with fat-tailed  degree distribution. We exploit this new implementation to study properties of the non-equilibrium
steady-state. 
We extend the dynamical cavity approach to calculate the pairwise correlations induced by different motifs in the network. 
Our results suggest that just two basic motifs of the network are able to accurately describe the entire statistics of observed correlations. Finally,  we investigate  models defined on networks containing bi-directional interactions. We observe that the stationary state associated  with networks with symmetric or anti-symmetric interactions  is biased towards the active or inactive state respectively, even if independent interaction entries are drawn from a symmetric distribution. This phenomenon, which can be regarded as a form of spontaneous symmetry-breaking, is peculiar to systems formulated in terms of Boolean variables, as opposed to Ising spins.

\end{abstract}
\maketitle

\section{Introduction}

A broad range of disordered and complex systems can be 
modelled in terms of binary units, which 
update their state 
according to stochastic Boolean functions of the neighbouring units.
They include spin glasses \cite{EdwardsAnderson75, SK75, viana1985phase} in physics, gene-regulatory and immune networks \cite{kauffman1969metabolic, derrida1986random, ParisiPNAS1990, AgliariEtAl2012, agliari2013immune} in biology, artificial neural networks \cite{Hopfield82, hertz2018introduction, Amit87a, sollich2014extensive}
in computer science, agent-based models \cite{ChalletZhang97, CoolenMG05, Iori99, Bornh01}, models of operational or credit risk \cite{ AnandKuehn07, HaKu06} in economics and finance, and a variety of hard combinatorial optimization problems \cite{weigt2000number, cocco2001statistical, martin2001statistical, Franz+01, mezard2002analytic} to name but a few. 
In recent years there was, in particular, a surge of interest in systems where each unit interacts only with a {\it finite} set of neighbours, and which are thus defined on finitely connected networks.

Statistical mechanics provides a rich arsenal of tools to analyse the equilibrium behaviour of sparse systems \cite{mezard2001bethe, viana1985phase,coolen2016replica,Annibale+10}.
However, the assumption of equilibrium is not satisfied in systems that are driven, dissipative or exhibit a degree of asymmetry in the interactions. Here, one has to resort to tools of non-equilibrium statistical mechanics, which are however much more challenging. 
Non-equilibrium dynamics has been analysed in explicit detail for a class of models where the dynamics of units has an absorbing state \cite{altarelli2013optimizing, lokhov2015dynamic, PagKu15,li2021impact} often combined with the simplifying features that interactions are of pure 
product form, e.g. in contagion processes \cite{KarrerNewm10,lokhov2014inferring}.

However, the dynamics of a large class of Boolean  models does not have an absorbing state. 
In the present manuscript, we analyse the stochastic dynamics of linear threshold models, formulated in terms of $\lbrace 0,1\rbrace$ variables, and discuss extensions to threshold models with multi-node interactions and  models with
Ising spin $\{\pm 1\}$, 
which do not have absorbing states. 
In such models, approximation schemes that are successfully used for dense systems, such as the heterogeneous  mean-field and TAP approaches \cite{mezard2002analytic,roudi2011dynamical}, have been shown to be ineffective for sparse systems \cite{zhang2012inference,aurell2012dynamic}. On the other hand, generating functional analyses  \cite{Dominicis78} can accurately characterise site averaged quantities, such as 
global magnetization and time-lagged correlations \cite{Hatchett+04,mimura2009parallel}, in sparse heterogeneous systems. However they require averaging over graph ensembles, hence they are not able to describe single instances. These can instead be investigated by the 
dynamic cavity method  \cite{mimura2009parallel,neri2009cavity, PagKu15, lokhov2014inferring} which is particularly effective in the analysis of sparse systems, whenever short loops are rare. In particular, the dynamic cavity method can potentially investigate dynamic properties at the level of individual nodes \cite{neri2009cavity}. Single-node statistics has, for instance, been shown to be highly heterogeneous for non-equilibrium models with an absorbing state \cite{lokhov2014inferring,KuRog17, altarelli2013optimizing}. Little, however, is known for other models, including spin models and linear threshold models.

In particular, explicit analysis has been feasible only for quite a narrow choice of networks, namely directed trees and fully asymmetric random graphs, where bi-directional links are sampled independently \cite{mimura2009parallel,neri2009cavity}. 
%
%
Moreover, 
both the dynamic cavity  and generating functional methods require evaluating averages over a configuration space that grows exponentially with the in-degree of the nodes, which limits the  applicability of both methods to networks with small in-degrees.
Due to this computational complexity barrier, a large class of networks, in particular those characterised by a fat-tailed degree distribution, are out of reach for analytical characterisation. Networks with fat-tailed degree distribution are, however, known to appear in many real-world problems and represent the outcome of generating network models based on preferential attachment
\cite{Lesk14, AlbBarab02, DorogMend03, NewBk10, Estrada2011, Latora+2017}. They are characterised by large differences in the local environment surrounding nodes, which 
are known to lead to
heterogeneous node properties at equilibrium \cite{Annibale+10}. Therefore, the aforementioned exponential complexity in the in-degrees has so far prevented a detailed study of node heterogeneities, precisely in those systems where they play a bigger role. 

We have recently proposed a method  based on dynamic programming  that overcomes this complexity barrier 
when
couplings are  chosen randomly from \textit{any} discrete set of equidistant values, 
the simplest example being couplings chosen from the set $\lbrace \pm J \rbrace$, see Ref.\, \cite{torrisi2021overcoming}. 
Our algorithm  reduces  the computational complexity in the in-degrees of the system from exponential to quadratic, thus allowing us to characterise the probabilistic evolution of individual node states in networks with a broad in-degree distribution.

In this manuscript, we extend our analysis of heterogeneous properties of the linear threshold model in fully asymmetric networks to 
include 
pairwise correlations between node states. Our analysis shows that the statistics of pairwise correlation is almost completely captured in terms of  simple motifs of length two. This reveals the extent to which, in terms of path distance, node states significantly influence each other. 

In addition, we extend our formalism 
to  investigate systems with multi-node interactions, using bipartite (or factor) graphs, which provide a more flexible model for combinatorial control; see Ref.\,\cite{fink2021boolean}. Multi-node interactions models can represent generic Boolean functions, which are of particular interest in the context of gene regulation \cite{kauffman1969metabolic,gates2021effective,Torrisi+20,Hannam+19}. 
We present our extended formalism for models of indicator variables $0,1$ as well as models of Ising spins $\pm 1$, for which our formulation in terms of bipartite graphs leads to (sparse) 
mixed $p$-spin models.

Finally, we  extend our analysis to  systems with bi-directional links.
For this class of systems, the cavity method
requires following the probability of the full dynamic trajectory, which cannot be reduced to the product of one-time-step terms, because  bi-directional links introduce retarded self-interactions of nodes with their past. 
This leads to an exponential scaling with the time horizon considered, which effectively  restricts 
the application of the method to  follow the dynamics over a few time-steps only, thus making the investigation of long time behaviour  infeasible.  An approximation technique called  the one-time approximation (OTA) \cite{neri2009cavity,aurell2012dynamic,zhang2012inference} has been proposed to overcome these restrictions.

The OTA factorises the probabilities of dynamic trajectories into one-time-step probabilities, and it leads to two sets of coupled equations, one for the marginals and one for the cavity marginals, which become closed under further assumptions. 
This makes the problem solvable also for the stationary state.
Both equations are affected by the aforementioned in-degree complexity, hence the OTA greatly benefits from our 
dynamic programming method. 
Within the OTA framework, different closure schemes have been proposed in the literature \cite{aurell2012dynamic,zhang2012inference}. We  evaluate the impact that different closure schemes have on the results, and we use such analysis to motivate our choice.  
A thorough inspection of the result leads to a surprising finding: networks with the same number of positive and negative interactions may sustain a biased distribution of node activation in presence of bi-directional links. This phenomenon, which can be regarded as a form of spontaneous symmetry breaking, is peculiar of systems with $\lbrace 0,1\rbrace$ representation of states (as opposed to Ising spins with $\lbrace \pm 1\rbrace$ representation).

Our paper is organized as follows. In Sec.\,2 we describe the dynamic programming approach to the linear threshold model on fully asymmetric networks with fat-tailed degree distribution.
Its non-equilibrium stationary state is 
characterised in terms of individual nodes statistics 
in Sec.\,3, and pairwise site correlations in Sec.\,4. 
In Sec.\,5 we extend our method both to models with multi-node interactions, formulated in terms of bipartite networks, and to networks of Ising spins. In Sec.\,6 we include bi-directional links in the analysis of linear threshold models, 
using the one-time approximation. In Sec.\,7 we investigate the unexpected occurrence of a form of spontaneous symmetry breaking in these systems. We summarise and discuss our results in Sec.\,8. Finally, we include two appendices: the first containing the high noise limit for the asymmetric network model and our multi-node interactions model, the second containing 
details concerning  the derivation of OTA schemes. 
The interested reader can find the code to reproduce the results shown in this paper at the following link. \footnote{ \url{https://doi.org/10.5281/zenodo.5996772}}

\section{Derivation of  dynamic cavity equation on fully asymmetric networks}
\label{sec:dynamical cavity}
We consider a directed network consisting of $N$ nodes labelled $i=1,\dots N$. For each edge $(i,j)$, the edge weight $J_{ij}\in \mathbb{R}$ defines the strength of the interaction carried from node $j$ to node $i$, while $J_{ij}=0$ if the link $(i,j)$ does not exist. In this section we will consider fully asymmetric networks, i.e. 
directed networks such that $\forall i,j$ if link $(i,j)$ exists, the link in the opposite direction $(j,i)$ does not exist. We  denote $\partial_i=\lbrace j: J_{ij}\neq 0 \rbrace$ the set of predecessors of node $i$ and $k_i^{\mathrm{in}}= |\partial_i|$  the in-degree of node  $i$. We will drop the superscript ``$\mathrm{in}$'' in $k_i^{\mathrm{in}}$ whenever it does not lead to ambiguity.   Every node $i$ can be in one of two states, described by a binary state variable  $n_i\in\lbrace 0,1\rbrace$  with $i \in \lbrace 1,\dots N \rbrace$. Several Boolean models can be defined on a given graph depending on the choice of the update rule of  node states, a prominent example being the random Boolean network model \cite{kauffman1969metabolic}.

In  the present section, we consider the linear threshold model defined on a weighted directed  network, which is specified by the interaction matrix $\bm{J} = (J_{ij})$ and thresholds $\lbrace \vartheta_i\rbrace_{i=1}^N$. A node $i$ evaluates a local field  
\begin{equation}
    h_i(\bm{n}_{\partial_i}) \defeq \sum_j J_{ij}n_j\,,
\end{equation} 
and becomes active in the next time-step if the sum of the local field and a random noise contribution $z_i$ exceeds the threshold $\vartheta_i$, i.e.,
\begin{equation}
    n_i(t+1)=\Theta \left(h_i\left( \bm{n}(t)\right) -\vartheta_i -z_i(t)\right),
    \label{eq: model}
\end{equation}
where the $z_i(t)$ are independent identically distributed random variables, 
with cumulative distribution  function  $\mathrm{Prob}[z_i(t)<x]\defeq\Phi_\beta(x)
$, for all $i, t$, and $T = \beta^{-1}$ is a parameter that characterises the noise strength.  
The local field $h_i(\bm{n}_{\partial_i}) $ depends on the states of nodes that are predecessors of node $i$.  We use $\bm{n}_{\partial_i}=\lbrace n_j ,j \in{\partial_i}\rbrace$ to denote the states of this set. In the following, we explicitly write the argument of the local field using the notation $h_i(\bm{n}_{\partial_i})$. Given the states $\bm{n}_{\partial_i}$, the probability that node $i$ is active at the next time-step is $\Phi_\beta( h_i(\bm{n}_{\partial_i})-\vartheta_i)$, the transition probability  
to state $n_i$, for node $i$,
is 
\begin{equation}
W[n_i|h_i(\bm{n}_{\partial_i}),\vartheta_i]= 
n_i\Phi_\beta( h_i(\bm{n}_{\partial_i})-\vartheta_i) +(1-n_i)(1-\Phi_\beta( h_i(\bm{n}_{\partial_i})-\vartheta_i) )\,.
\label{eq:conditional_step}
\end{equation}
Given the joint probability $P(\bm{n}_{\partial_i},t)$ of the states $\bm{n}_{\partial_i}$ at time t, the probability  that node $i$ is in state $n_i$ at time $t+1$ is given by
\begin{equation}
P(n_i,t+1)=\sum_{\bm{n}_{\partial_i}}W[n_i|h_i(\bm{n}_{\partial_i}),\vartheta_i]P(\bm{n}_{\partial_i},t)\,, 
\label{eq: dynamical magnetisation}
\end{equation}
where  
$\sum_{\bm{n}_{\partial_i}}(\cdot)\defeq\prod_{j \in\partial_i} \sum_{n_j=\lbrace 0,1\rbrace} (\cdot)$ indicates the sum over all possible configurations of the microscopic variables. 
For finitely coordinated random graphs of the type considered here, the cavity method \cite{mezard2001bethe} can be used to analyse the dynamics of the system. It is exact on trees and known to become exact for finitely coordinated random graphs in the thermodynamic limit, as the typical length of any loops in such systems diverges logarithmically in system size $N$. 
On the cavity graph where node $i$, along with the edges connected to it, has been removed, the cavity approximation assumes 
\begin{equation}
P^{(i)}(\bm{n}_{\partial_i},t)=\prod_{j \in \partial _i}P^{(i)}(n_j,t)\,,
\end{equation}
which holds on a tree-like structure.
The probability
$P^{(i)}(\bm{n}_{\partial_i},t)$  of the states of nodes neighbouring $i$ will in general be different from 
the corresponding probability on the original graph, i.e. $P(\bm{n}_{\partial_i},t)\neq P^{(i)}(\bm{n}_{\partial_i},t)$, but equality does hold on a directed tree and it holds approximately on a random fully asymmetric graph, in which loops are rare  \cite{mimura2009parallel,neri2009cavity,Torrisi+20}\footnote{Cavity method can be applied also in the general case, as  discussed in \cite{neri2009cavity} and in \ref{sec:derivation OTA}. However the evaluation is more challenging.}.  
 Using the equivalence of the cavity and non-cavity dynamics, from \Eref{eq: dynamical magnetisation}
 \begin{equation}
P(n_i,t+1)=\sum_{\bm{n}_{\partial_i}} W[n_i|h_i(\bm{n}_{\partial_i}),\vartheta_i] \prod_{j \in \partial _i}P(n_j,t)\,.
\label{eq:cavity_0}
\end{equation}
In particular, the probability of node activation becomes
\begin{equation}
P(n_i=1,t+1)=\sum_{\bm{n}_{\partial_i}} \Phi_\beta  \left( h_i(\bm{n}_{\partial_i})-\vartheta_i\right) \prod_{j \in \partial _i}P(n_j,t)\,.
\label{eq:cavity}
\end{equation}
Let us denote by $P_i(t) 
\defeq P(n_i=1,t)$ the activation probability of node $i$ at time $t$, let the angle bracket $\langle\,\cdot\,\rangle_{ \bm{n}_{\partial_i},\,t}$ indicates an average over the states of the predecessors of node $i$, evaluated using their joint node activation probabilities at time $t$
\begin{equation}
\langle(~\cdot~)\rangle_{ \bm{n}_{\partial_i},\,t}\defeq\sum_{ \bm{n}_{\partial_i}} \prod_{j\in\partial_i}P_j(t)^{n_j}[1-P_j(t)]^{1-n_j}\,(~\cdot~)\,.
\label{eq:average_definition}
\end{equation}
Combining \Eref{eq:average_definition} and \Eref{eq:cavity} a concise expression for the activation probability $P_i(t+1)$ is obtained as 
\begin{equation} P_i(t+1)=
\left\langle \Phi_\beta\left(h_i(\bm{n}_{\partial_i})-\vartheta_i \right)\right\rangle_{\bm{n}_{\partial_i},\,t}\ .
\label{eq: result cavity}
\end{equation}
The average in \Eref{eq: result cavity} is evaluated over a configuration space containing $2^{k_i}$ elements, as shown in \Eref{eq:average_definition}, thus the evaluation of the average hits an (exponential) complexity barrier whenever $k_i\gg1$.  In Ref.\,\cite{torrisi2021overcoming} we have proposed a new method that  uses dynamic programming to efficiently perform the average in \Eref{eq: result cavity} using a polynomial number of operations. In the following, we recap the main ideas of the method.
\paragraph{Dynamic programming}
 \Eref{eq: result cavity} depends on the distribution  of states $\mathbf{n}_{\partial_i}$ through the local field $h_i(\mathbf{n}_{\partial_i})$. If couplings are extracted from a discrete set of values, e.g., $J_{ij}=\lbrace 0,\pm J \rbrace$, there is a degeneracy of values of $h_i(\bm{n}_{\partial_i})$, i.e., several configurations of $k_i$ states $(n_{j_1},\dots n_{j_{k_i}})$ are associated with the same value of local field. In fact, for a given realisation of the interaction terms $\left(J_{ij_1}\dots J_{ij_{k_i}}\right)$ the number of possible values of  $h_i(\bm{n}_{\partial_i})$ is $k_i+1$, while the number of configurations of states is $2^{k_i}$. We exploit this degeneracy by averaging directly over the local fields, rather than over the micro-states of the set of predecessors, using a  dynamic programming approach. We show below our approach   reduces the computational complexity of the problem from $\mathcal{O}(2^{k_i})$ to $\mathcal{O}(k_i^2)$, \cite{torrisi2021overcoming}.  

Let us consider a node $i$ with in-degree $k_i$, and let us use $\lbrace 1,\dots k_i\rbrace$ to label the indices of the predecessors of node $i$.
We first define the sub-problem
\begin{equation}
      f_i(\ell,\tilde{h}) = \left\langle\! \Phi_{\beta}\Bigg(\tilde{h}+\sum_{j=\ell}^{k_i}J_{ij}n_j -\vartheta_i\Bigg) \right\rangle_{n_{\ell,\dots,k_i}, t}\, \mbox{for }\ell \in \lbrace 1,\dots k_i\rbrace,
  \label{eq:dynamic_programming_f_def}
\end{equation}
which consists in performing the average over a sub-set of nodes. We will refer to $\tilde h$ as an auxiliary field.
At any given $\ell$, the set of $f_i(\ell,\tilde{h})$ of interest would in fact correspond to the set of averages representing the sub-problems of \Eref{eq: result cavity} that remain to be evaluated after the average over the first  $\ell-1$ nodes has been performed.
Quantities in \Eref{eq:dynamic_programming_f_def} are not performed directly, but through the recursive relationship
\begin{equation}
   f_i(\ell,\tilde{h}) = P_\ell(t)\, f_i(\ell+1,\tilde{h} +J_{i\ell})
   +\big(1-P_\ell(t)\big)\,f_i(\ell+1,\tilde{h}) 
   \label{eq: recursive}
\end{equation}
for $1\leq \ell\leq k_i$, with the {\em terminal boundary condition}
\begin{equation}
  f_i(k_i+1,\tilde{h}) = \Phi_{\beta}\left(\tilde{h} -\vartheta_i\right)\ .
\end{equation}
The original average $P_i(t+1)$ of  \Eref{eq: result cavity} that we are ultimately interested in is, within this backward iteration scheme, obtained as
\begin{equation}
    P_i(t+1) = f_i(1,0)\ .
\end{equation}

In our earlier work, Ref.\,\cite{torrisi2021overcoming}, we have shown that \Eref{eq: recursive} leads to a dramatic reduction of the computational complexity 
for binary interactions $J_{ij}\in\{0, \pm J\}$.
The computational and memory requirement for the evaluation of $P_i(t+1)$ is seen to scale as $k_i^2$, for a node with in-degree $k_i$, rather than $2^{k_i}$, as it would in a naive evaluation.
 This entails that the complexity of our algorithm is $\mathcal{O}{(\sum_i k_i^2)}$.  
\section{Node activation heterogeneity}
\label{sec:node_heterogeneity}
It has been shown that  fully asymmetric graphs do not exhibit a spin-glass phase  in a variety of dynamical models, and the time-dependent local magnetizations or node activation probabilities are expected to converge to a fixed point \cite{hertz1986memory,gutfreund1988nature,Hatchett+04,neri2009cavity}. We exploit this fact to investigate the trajectory of nodes' states by iteration of \Eref{eq: result cavity}.
Our dynamic programming implementation provides an efficient method to evaluate  node activation probabilities. We apply our formalism to inspect the dynamical properties of networks with fat-tailed degree distributions. The theoretical evaluations presented in this section  would not be feasible without our dynamic programming implementation.
\paragraph{Model Parameters Used in this Study}
In the present section, we consider models of synthetic  networks in the configuration model class, i.e. we will study ensembles of maximally random directed graphs ${\bf J}$ with constrained in-degree  and out-degree sequences of nodes. The in-degree of nodes are drawn from the distribution $\rho^{\mathrm{in}}(k)
= N^{-1}\sum_i \delta_{k,k_i^{\mathrm{in}}}$. In this paper, we use the same distribution for the  out-degree of nodes $k_i^{\mathrm{out}}$.\footnote{If the distributions of -in- and out-degree are drown from different distributions, a minimal requirement is  $\sum_i k_i^{\mathrm{in}}=\sum_i k_i^{\mathrm{out}}$ that guarantees the conservation of the number of links.} The in-degree and out-degree for the same node are drawn independently.   We use $\langle k^{\mathrm{in}}\rangle$ to denote the mean in-degree. Our synthetic model uses parameters observed in real-world networks.
 Specifically, we use parameters extracted from a gene regulatory network \cite{han2018trrust}, which is characterised by a node in-degree distribution  with a power-law tail $P_\gamma(k)\sim \gamma k^{-\gamma-1}$ and $\gamma=2.81$, which we implement in terms of the discrete fat-tailed distribution $P_\gamma(k) = k^{-\gamma}-(k+1)^{-\gamma}$; it is defined for positive integers $1\le k\in \mathbb{N}$ and has the desired power-law behaviour for large $k$. 
For every link  $(i,j)$, the corresponding interaction term $J_{ij}$ is sampled from the distribution $P(J_{ij})=\eta \delta(J_{ij}-J) + (1-\eta)\delta(J_{ij}+J)$ and we use $\eta = 0.621$ as suggested by the data in \cite{han2018trrust} and $J = 1/\sqrt{\langle k^{\mathrm{in}} \rangle}\approx 0.72$ unless stated otherwise. We set the value of the threshold to be the same  for all sites, $\vartheta_i=\vartheta$ $\forall i$.
We assume a ``thermal'' noise distribution with $\Phi_\beta(x) = 1/2\left(1+\tanh\frac{\beta x}{2}\right)$.  

\subsection{Comparison between theory and simulation }
\label{sect:comparison}
\begin{figure}
\begin{subfigure}{0.5\textwidth}
\includegraphics[width = \textwidth]{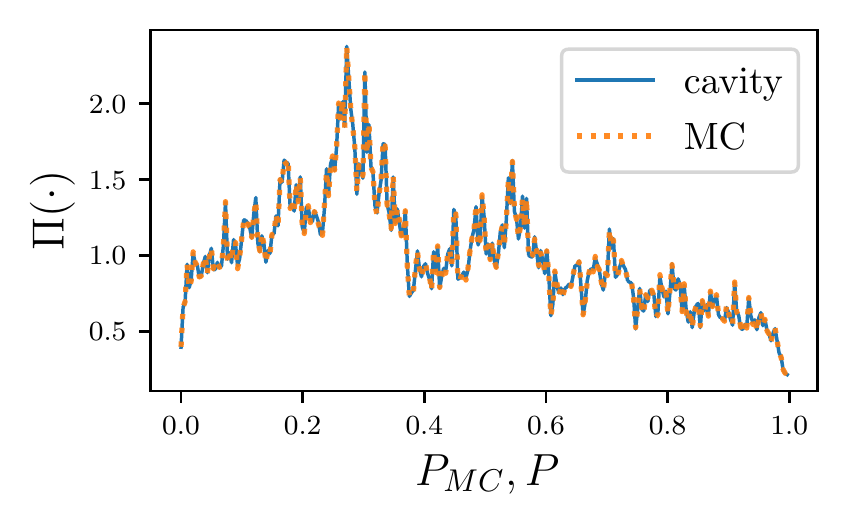}
\end{subfigure}
\begin{subfigure}{0.5\textwidth}
\includegraphics[width = \textwidth]{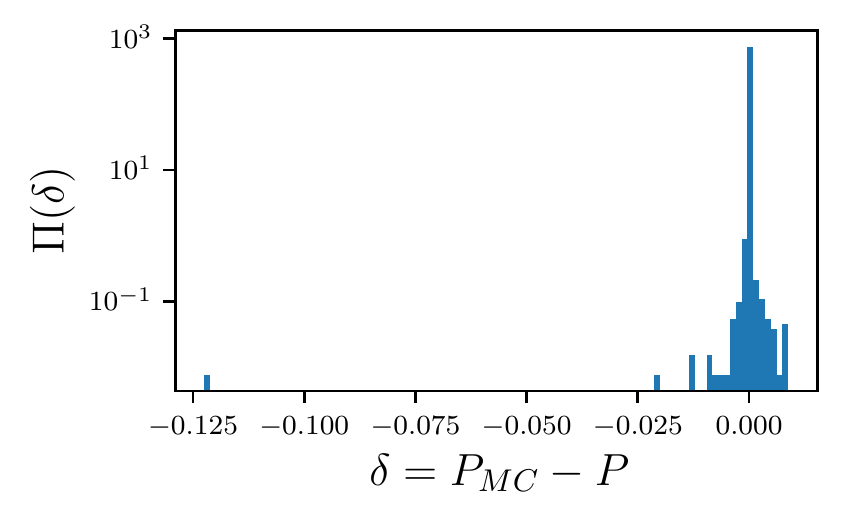}
\end{subfigure}
\caption{Left: distribution of the node activation probability, as computed from cavity theoretical calculation (blue solid line) and from Monte Carlo direct simulation of the microscopic dynamics (dotted green line). Right: distribution of the differences $\delta_i =P_{i,MC}- P_{i}$, evaluated for each node $i$ of the network.  Parameters: $N =100,000, T/J=0.2,\gamma=2.81,\eta=0.5,\vartheta/J = 0.1,J = 1$. }
\label{fig:cav_vs_sim}
\end{figure}

The dynamic programming algorithm allows to speed up the evaluation of the node activation probability at time $t+1$ given the knowledge of node activation probabilities at time $t$. By iterating the procedure over the time-steps of interest, we solve the dynamics for the nodes activation probabilities. In this section, we compare the analytical results against Monte Carlo simulations of the microscopic dynamics. Several results in the literature indicate that the dynamic cavity method is effective to reproduce the  trajectory of fully asymmetric graphs, see Refs.\, \cite{neri2009cavity,del2015dynamic}.  While those results inspect  the site average of node activations for  Ising spins, we consider  the full distribution of node activations for indicator variables.  In particular, we compare the stationary values of the node activation probabilities $P_i = \lim_{t\rightarrow \infty}P_i(t)$ $\forall i \in\lbrace 1,\dots N\rbrace $ obtained from simulations with those obtained via the cavity approach. The  stationary activation probabilities are obtained by running the iterative procedure \Eref{eq: result cavity}  until convergence. The convergence  is controlled by measuring  the difference between activation probabilities $\epsilon_{t+1}=\mathrm{max}_i |P_i(t+1)-P_i(t)|$ in two consecutive steps;  this criterion is adopted in the entire  manuscript. Let us call $P_{i}=P_i(t)$ corresponding to the smallest $t$ satisfying the exit condition, which we took to be $\epsilon_t<10^{-4}$.  
We test cavity predictions  against  simulations of the microscopic dynamics \Eref{eq: model}. 
 We evaluate the stationary node activation probability of simulated dynamical trajectories by taking the sample average of node activations
 \begin{equation}
P_{i,MC}= \frac{1}{t_s}\sum_{t={t_0}}^{t_0+t_{s}}n_i(t)\,,
\label{eq:sim_trj}
\end{equation}
for $t_{s}\gg 1$, and $t_0$ a sufficiently long time to allow the dynamics to relax to stationarity.  In the following, we use the general notation $\Pi(x)$ to denote the probability density function for the realisation of local quantities $x_i$, \textit{e.g.} $\Pi(P)= N^{-1}\sum_i\delta(P_i-P)$. In Fig.\,\ref{fig:cav_vs_sim} we show the distribution of the node activation probability  as computed via the cavity method and by estimation from simulations. These are found to be in excellent agreement, confirming the validity of the cavity method to describe the stationary state, see Fig.\,\ref{fig:cav_vs_sim}.  We note that,  to reach the resolution imposed by cavity $\epsilon_t<10^{-4}$ through simulations, we need to simulate long trajectories in \Eref{eq:sim_trj} with $t_s\sim10^{8}$, which makes the evaluation from simulation computationally far more expensive than an evaluation based on the cavity method. For the same realisation of a network of $N= 100,000$ nodes and identical parameter choices, the cavity implementation took \SI{9}{\second}, whereas simulations took more than \SI{11} days on the same machine.

\paragraph{Time evolution }

\begin{figure}
\centering
   \includegraphics[width= 0.6\textwidth]{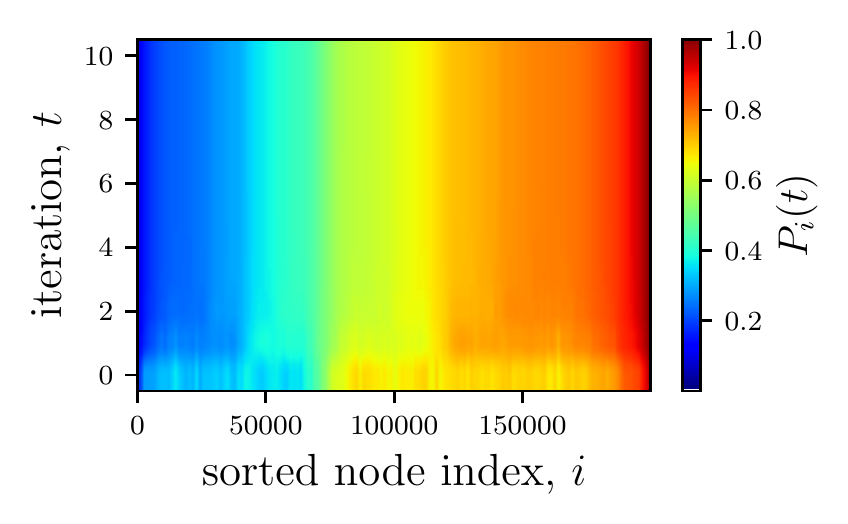}
         \caption{
         Trajectory of probabilities  of node activation up to convergence. The colour of the pixel located at row $t$  and column $i$ maps  the value $ P_i(t)$.   Colour-bar on the right maps the colours to the value of activation probability.  For visualisation purposes nodes  are sorted by the value of  $P$ at the end of iteration, \textit{i.e.}, column locations are sorted according to their value in the top row.  Parameters: $N=200000$, $\vartheta/J = 0.2$, $T/J= 0.3, J \approx 0.72$, $\eta = 0.621$.}
        \label{fig:trajectory}
\end{figure}

\begin{figure}
\centering
\includegraphics[width =\textwidth]{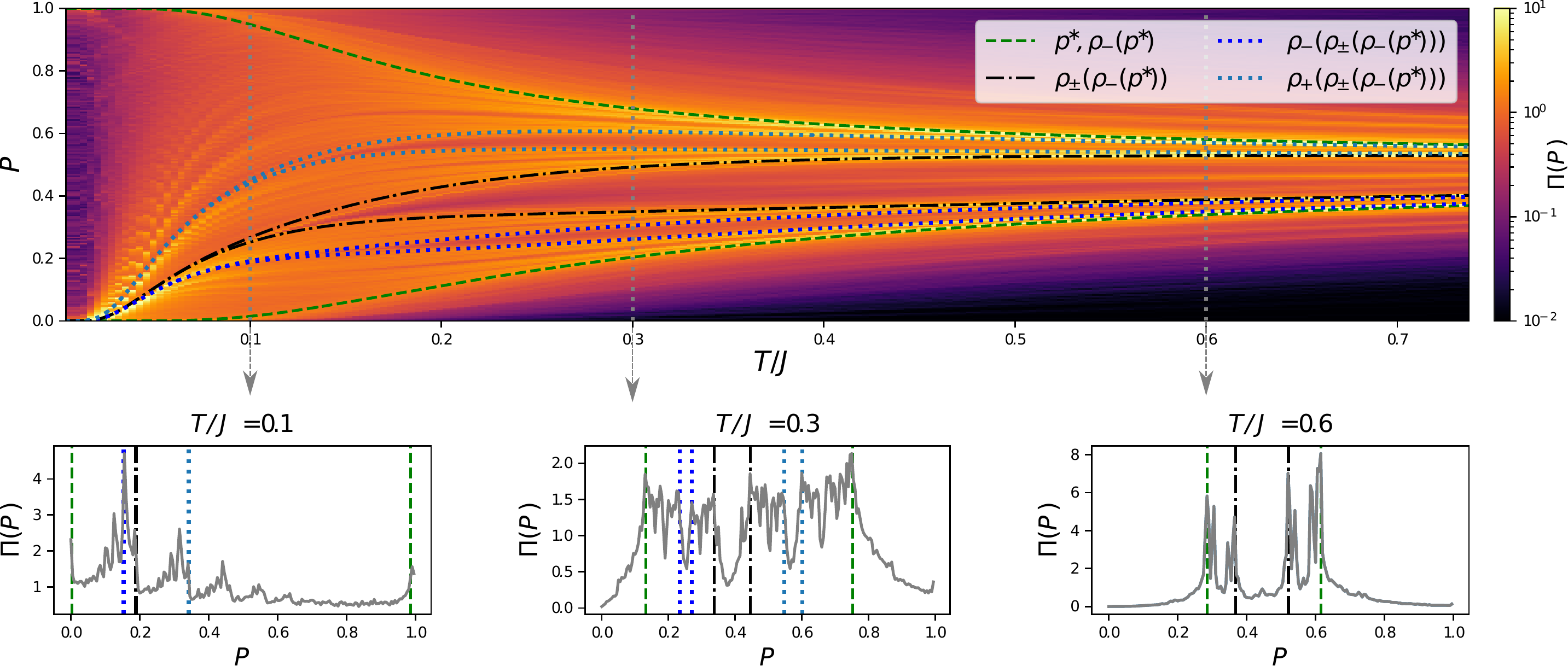}
\caption{Heat-map of the distribution of  node activation probabilities  $\Pi(P)$   at different values of noise parameter $T/J$ (top). 
Dashed, dotted,  and dot-dashed  lines show the graph of different function compositions of  $\rho_\pm$ at different values of $T$, see \Eref{eq:analytical_map} . Vertical dotted lines mark the noise levels at which the histograms of $\Pi(P)$ are shown in the lower  panels. They  probe the regimes at low, intermediate and high noise, corresponding to  $T/J=0.1, 0.3, 0.6$  from left to right. The central lower panel corresponds to the setting at the convergence of Fig.\, \ref{fig:trajectory}. Same parameters as in Fig.\, \ref{fig:trajectory} are adopted.}
\label{fig: power_law}
\end{figure}

 We  investigate the statistical properties of dynamical trajectories through forward iteration of \Eref{eq: result cavity}. We initialise the system with random initial conditions for the node activation probabilities $P_i(0)$.  We stop iterations at  convergence using the same criterion as above. 
 The time evolution of the dynamics for each site is shown in Fig.\,\ref{fig:trajectory}.  Node activation probabilities converge after very few time-steps. This property is linked to the absence of feedback signals in the dynamics, thanks to the directed tree-like structure of the network.
The probability density function of the node activation probabilities at stationarity is shown in Fig.\,\ref{fig: power_law}  for different  noise levels $T$. The distribution of the node activation probability is not uni-modal,  indicating the need to inspect the full distribution of node activation, as measures of  central tendency are not representative. This is also true in the high noise limit, where the problem is directly solvable; see \ref{sec:high noise}.  

\begin{figure}
 \centering
   \includegraphics[width= 0.7\textwidth]{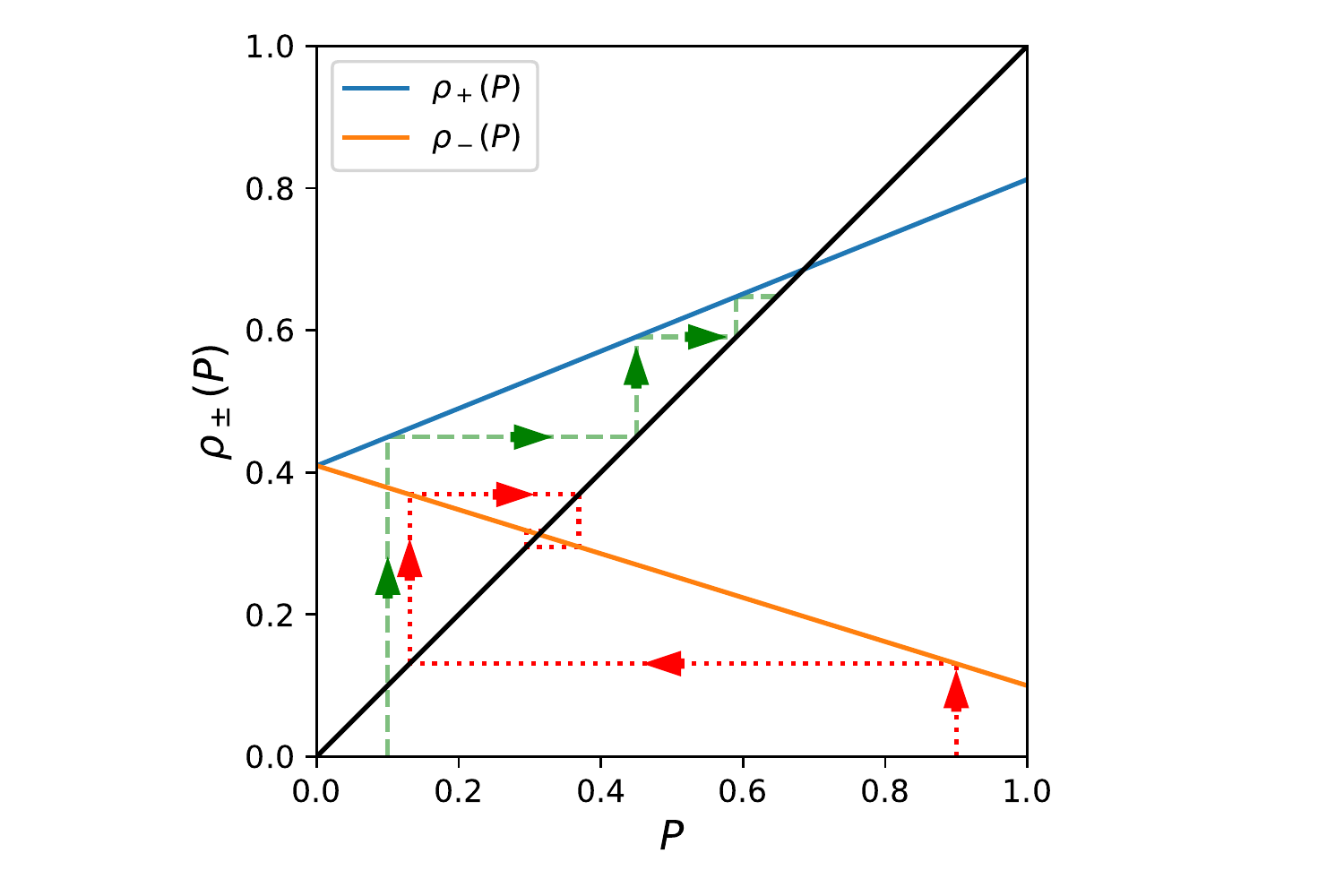}
   \caption{Graphical solution of the equation  $\rho_{\pm}(P^*)=P^*$. The black line represents the  graph of the function $f(x) = x$. Fixed points $p^*_\pm$ are obtained by the intersections of the lines $\rho_\pm$ with the bisector. Same parameters as in Fig.\,\ref{fig:trajectory}.}
   \label{fig:analytical}
 \end{figure}

\begin{figure}
\centering
   \includegraphics[width=0.6 \textwidth]{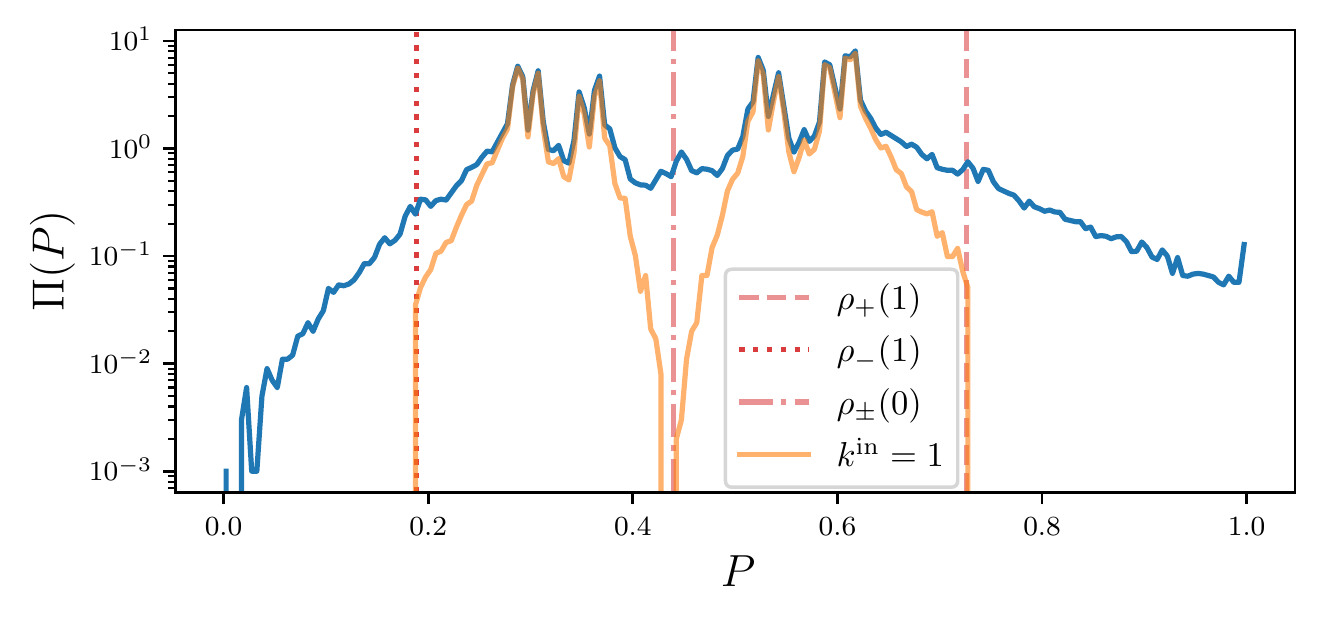}
   \caption{Distribution of the node activation probability and contribution from nodes with in-degree 1 (orange). Vertical lines indicated the range of possible values of $P$ for nodes with in-degree 1, as predicted by Fig.\, \ref{fig:analytical}. Same parameters as in Fig.\,\ref{fig:trajectory}.}
   \label{fig:degree1}
\end{figure}

The multi-modal structure of the node activation distribution can be rationalised by simple reasoning. Thanks to the fat tailed property of the network, more than \si{87\percent}  of nodes have in-degree 1.  Let us consider  a node $i$ with only one predecessor $j$,  let us call $\rho_\pm(P)$ the probability that node $i$ is active, given that $J_{ij}=\pm J$, and that the node activation probability of that predecessor is $P$, i.e., $\rho_\pm(P)=\left\langle\Phi_\beta(\pm Jn-\vartheta)\right\rangle_{n}$ with $\langle n\rangle_n = P$. 
The explicit expression of functions $\rho_\pm(P)$ becomes
\begin{equation}
\begin{split}
 \rho_\pm (P) &= \frac{P}{2}\left[1 + \tanh\frac{\beta}{2}\left(\pm J-\vartheta \right) \right] +
 \frac{1-P}{2}\left[1 -\tanh\frac{\beta\vartheta}{2}\right]\\
& =\frac{1}{2}\left[1-\tanh\frac{\beta\vartheta}{2}\right]+\frac{P}{2}\left[\tanh\frac{\beta}{2}\left(\pm J-\vartheta \right)+\tanh\frac{\beta\vartheta}{2} \right]   \,.
\end{split}
 \label{eq:analytical_map}
\end{equation}
  The properties of  nodes with in-degree 1 can be captured in terms of the map shown in Fig.\,\ref{fig:analytical}. We denote $p^*$ the fixed point defined as $p^* = \rho_+(p^*)$.  The composition of function $\rho_\pm$ is able to describe the trends observed in Fig.\,\ref{fig: power_law}. In Fig.\,\ref{fig:degree1}, we highlight the contribution to the  distribution of node activation probability from nodes with in-degree 1. The values of  $\rho_{\pm}(1)$ give the lower and upper bounds of the node activation probability for nodes with in-degree 1. Even though nodes with in-degree 1 account for the vast majority of nodes, the dynamical properties of node activation depends on the contribution of other nodes as well. 

\subsection{Comparison with the naive mean-field approximation}
Solving the dynamics  of the linear threshold model for generic couplings is computationally demanding as already discussed above.  
In the context of disordered systems, 
approximate methods have been proposed, 
which are effective for models with dense interactions, such as the dynamic naive mean-field (nMF) approximation and dynamic TAP equations, for interactions with an arbitrary degree of symmetry
\cite{roudi2011dynamical}, and specifically for 
fully asymmetric interactions
\cite{mezard2011exact}. 
However, for diluted systems, dynamic nMF approximations have been shown to be less effective \cite{gleeson2011high,aurell2012dynamic}
for systems with both ferromagnetic and spin-glass interactions, especially at low temperature. 
In this section, we investigate the distribution of node activation and we highlight that even in the region of  parameters where the site average of the node activation probability is well captured by the nMF method, the nMF turns out to be incapable of reliably capturing the full distribution of node activation probabilities. 

We briefly expose the intuitive idea of the mean-field approximation, and we refer to Ref.\,\cite{aurell2012dynamic} for more details.
For a node $i$ let us consider the local field $h_i=\sum_{j}J_{ij}n_j$ and the probability associated $\mathrm{Prob}(h_i)=\langle \delta(h_i-\sum_{j}J_{ij}n_j)\rangle_{\bm{n}}$.
The core idea of the mean-field approximation is to approximate $\mathrm{Prob}(h_i)$ with a delta-function peaked on the mean  value $\langle h_i \rangle$, which implies  $\left\langle \Phi_\beta\left(h_i-\vartheta_i \right)\right\rangle\approx  \Phi_\beta\left(\left\langle h_i\right\rangle-\vartheta_i \right)$. This assumption is motivated in the limit where the number of neighbours of node $i$  diverges where,
thanks to the law of large numbers,  one expects $\mathrm{Prob}(h_i)$ to concentrate --- for suitably normalized $J_{ij}$ --- around the mean value. However, in the case of sparse matrices, the number of neighbours is typically small, therefore,  we expect the nMF to fail. However, our results below indicate that the site averaged node activation is well captured by the mean-field approximation in a variety of conditions.

\begin{figure}
\centering
\begin{subfigure}{0.495\textwidth}
\includegraphics[width = \textwidth]{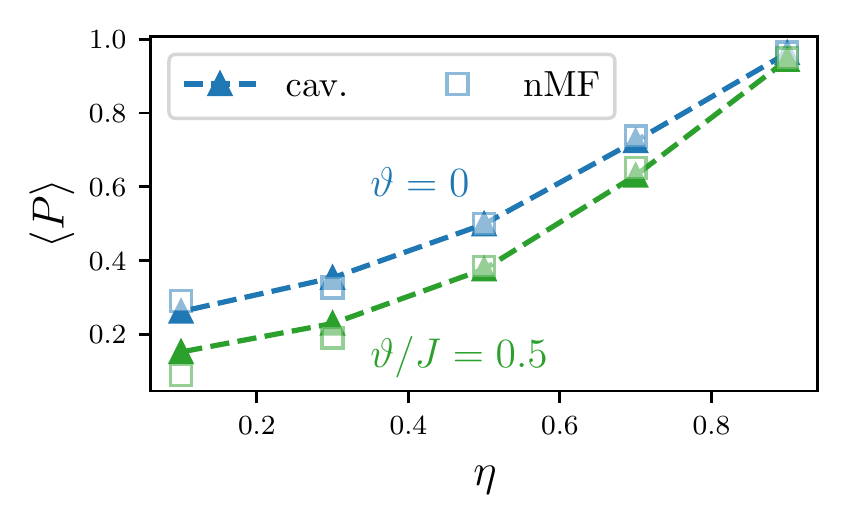}
\end{subfigure}
\begin{subfigure}{0.495\textwidth}
\includegraphics[width = \textwidth]{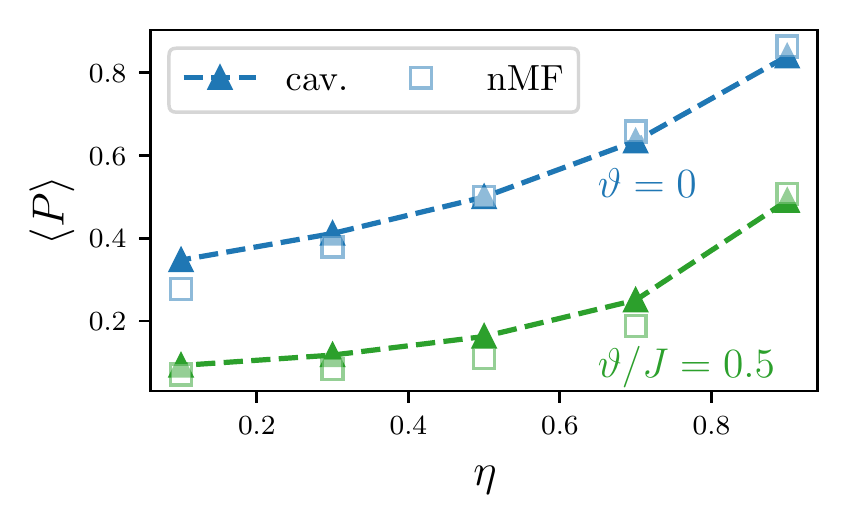}
\end{subfigure}
\begin{subfigure}{0.495\textwidth}
\includegraphics[width = \textwidth]{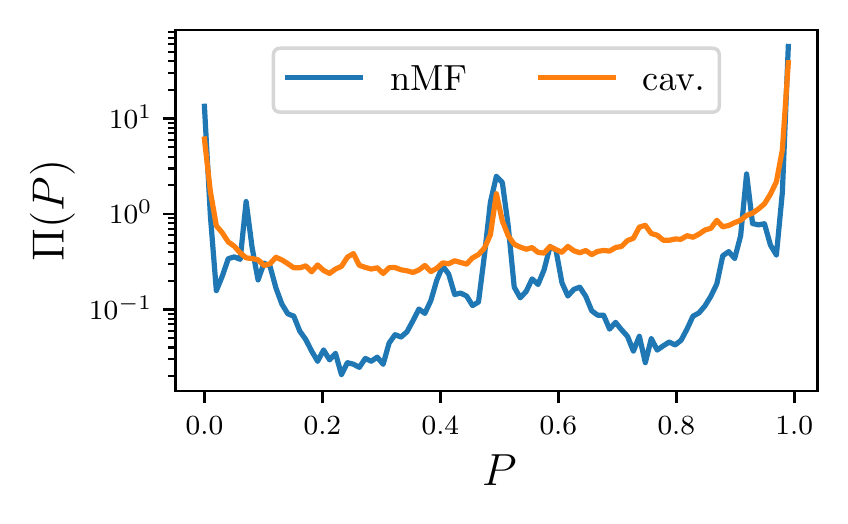}
\end{subfigure}
\begin{subfigure}{0.495\textwidth}
\includegraphics[width = \textwidth]{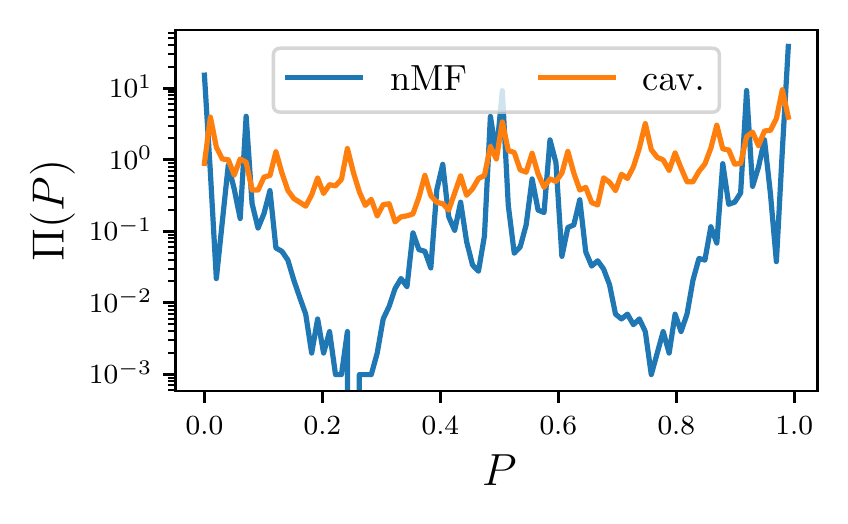}
\end{subfigure}
\caption{
Comparison between naive mean-field (nMF) and cavity (cav) solutions at $T/J=0.2$ in random directed 
networks with $N=100000$ nodes and regular topology with $k^{\mathrm{in}}=k^{\mathrm{out}}=3$ (left) and heterogeneous in- and out-degrees, drawn from a fat-tailed distribution with $\gamma=4$ (right). Upper panels show the average activation probability $\langle P\rangle=N^{-1}\sum_{i=i}^N P_i$ versus the bias $\eta$, for $\vartheta/J=0, 0.5$. Results from cavity are shown by triangle markers, while squares show mean-field results. 
Lower panels show the distribution $\Pi(P)$ of the site-dependent activation probabilities $\lbrace P_i \rbrace_{i=1}^N$, for $\vartheta=0$, $\eta=0.7$. Orange lines show results from cavity \Eref{eq:cavity}, blue lines show results from naive mean-field \Eref{eq:mean_field}.
}
\label{fig:mean_field}
\end{figure}

Starting from \Eref{eq: model}, the associated mean-field equation for fully asymmetric networks is
\begin{equation}
P_i(t+1) = \Phi_\beta \left(\sum_j J_{ij}P_j(t)-\vartheta_i\right)\,,
\label{eq:mean_field}
\end{equation}
which represents an approximation of \Eref{eq:cavity}. We run \Eref{eq:mean_field}  for each node $i$ starting from random initial conditions until convergence.  We apply the heterogeneous mean-field in the same class of networks as above (fully asymmetric adjacency matrix with  discrete coupling strength) and use our algorithm as ground truth  to evaluate the performance of the mean-field approximation. We compute the site average of the node activation probability $\langle P\rangle=N^{-1}\sum_iP_i$  both in the case where the set $\lbrace P_i\rbrace_{i=1}^N$ is computed by using \Eref{eq:mean_field} and    \Eref{eq:cavity}. In Fig.\,\ref{fig:mean_field} we compare the results for the site average of node activation  at different values of bias $\eta$ and threshold $\vartheta$ for two  network models, namely the directed random regular graphs, 
 and  networks with fat-tailed degree distribution. 
Results indicate the mean-field  captures remarkably well the behaviour of  the site  average for  different choices of   network models and both in the absence and presence of an  external threshold. However,  the mean-field approximation captures poorly the heterogeneous site dependence, and the distribution $\Pi(P)$ presents large differences  between the cavity and mean-field results. 

\section{Pairwise correlation for the linear threshold model on fully asymmetric networks}
\label{sec:correlation}
Temporal and pairwise correlations are of  great interest for the analysis of dynamics on networks. 
For example, correlations are used extensively to discover patterns of relationships in numerous inverse problems \cite{deplancke2012gene,bitbol2016inferring,morcos2011direct,cavagna2014dynamical}.
In the context of spin models on networks,  the correlation has been investigated in the dense regime for asymmetric interactions through mean-field  \cite{mezard2011exact} and an analytical expression for the  site correlators can be formally solved  through TAP equations for generic system   not at equilibrium \cite{roudi2011dynamical}. However,  in the latter case, the analytical equations are difficult to solve. In this section, we introduce a method to evaluate spatio-temporal correlations in  the sparse regime using dynamic cavity techniques.

As shown above, in the cavity approach  the evaluation of the one-point expectations $\langle n_i\rangle$   relies on the assumption that the states associated with the predecessors of a node are independent of one another, which is true on a tree.  
Instead, for a given {\em pair\/} of nodes, the states of the {\it joint} predecessors are not independent of one another, even on a tree. The cavity method has recently been applied in the presence of node correlations on loopy graphs  to investigate 
percolation, random matrix spectra  \cite{cantwell2019message}, and the equilibrium state of Ising spin models \cite{kirkley2021belief}. 
The authors explicitly include the contribution from short loops, e.g., triangles, which break the assumption of independence of states the cavity method relies on. Here, we solve  a different problem, namely, we adapt the cavity method to evaluate  two-points expectations. In our context,   a new set of motifs that break the assumption of independence of states arises, which  are present even in directed trees.   Therefore, we devise a method that incorporates node correlations inside the cavity equations. 

If the graph admits a  giant connected component, every set of nodes belonging to the giant out-component  shares at least one common ancestor, which  would make the evaluation of node correlation infeasible on a large graph. 
In the following, we assume correlations to be mainly generated by two network motifs, where nodes have common predecessors either at distance one or two, and we test our approximate results against direct simulations. 

\begin{figure}
\centering
\begin{subfigure}{0.48\textwidth}
\includegraphics[width = \textwidth]{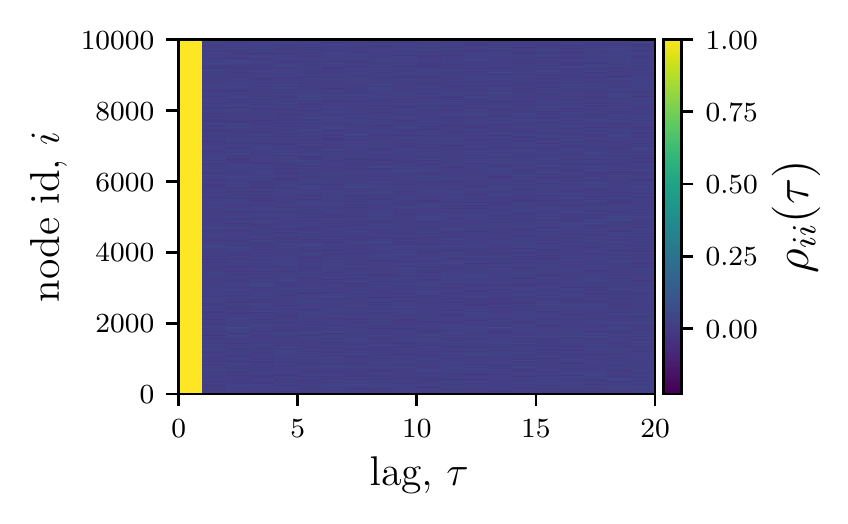}
\end{subfigure}
\begin{subfigure}{0.48\textwidth}
\includegraphics[width = \textwidth]{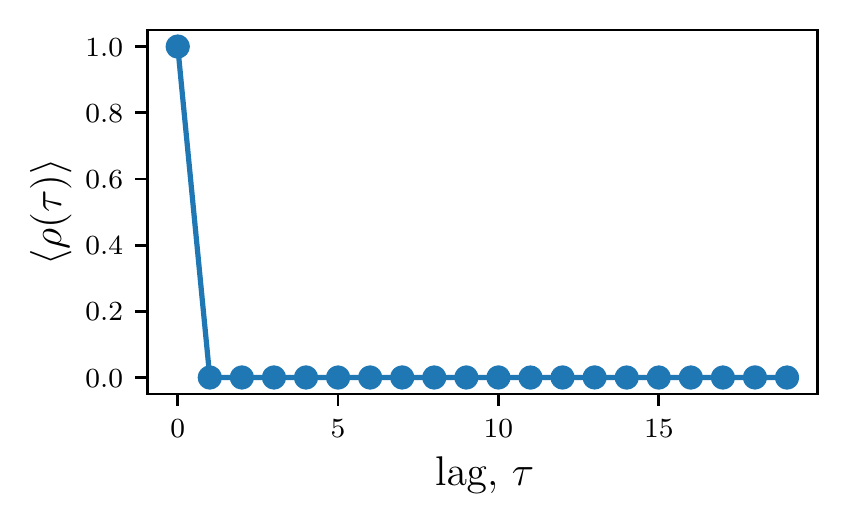}
\end{subfigure}
\caption{ Single-site lagged Pearson autocorrelation $ \rho_{ii}(\tau)$ for different nodes $i$ (left) and its site average $\langle \rho(\tau)\rangle$ (right) as a function of the time-lag $\tau$ in  a directed random regular graph with parameters: $N=10000$, $k^{\mathrm{out}}=k^{\mathrm{in}}=2$, $T/J = 0.2,\vartheta = 0$.  }
\label{fig:time_correlation}
\end{figure}

\begin{figure}
\begin{subfigure}{0.48\textwidth}
\centering
\includegraphics[width = 0.6\textwidth]{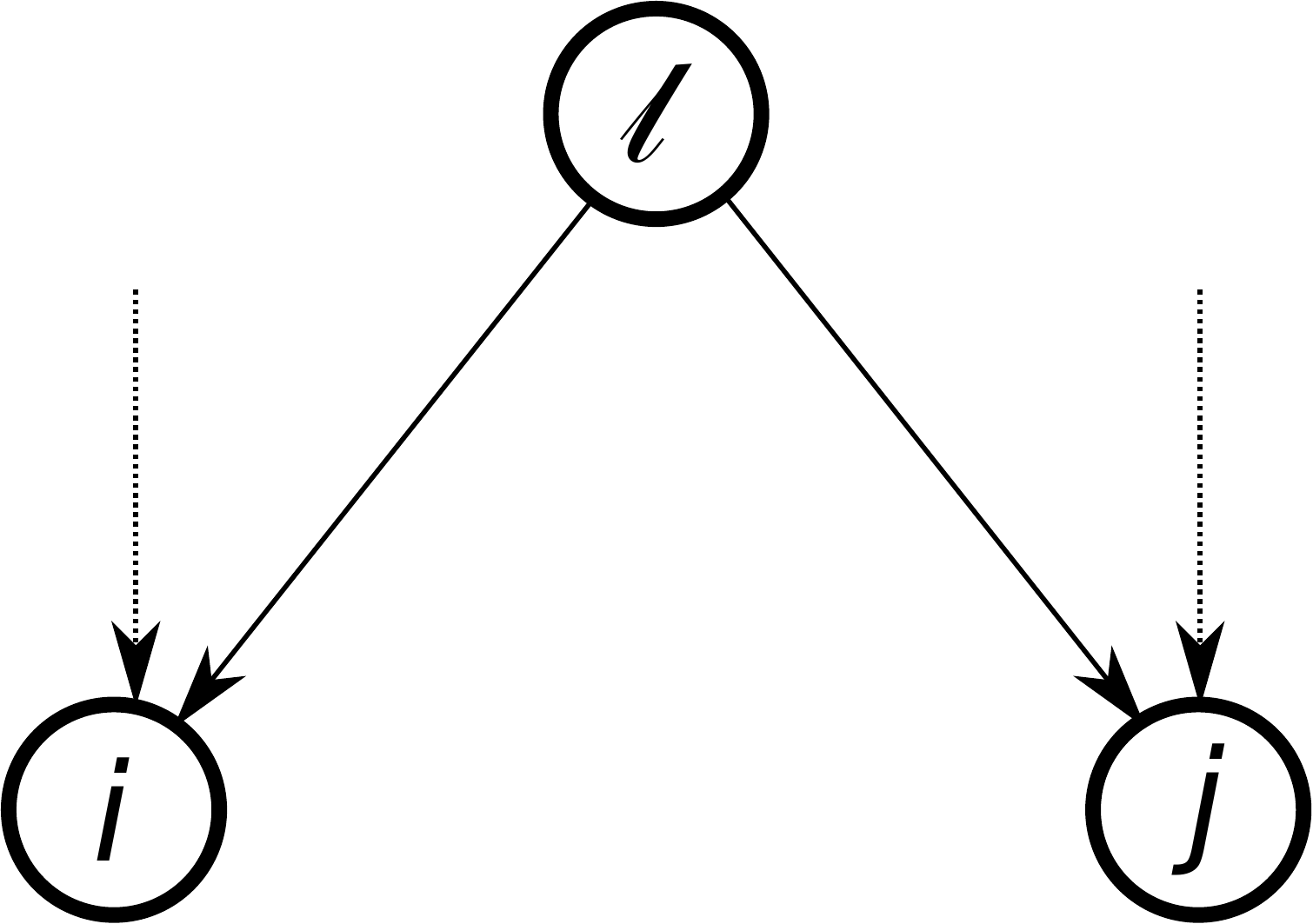}
\caption{}
\end{subfigure}
\begin{subfigure}{0.48\textwidth}
\centering
\includegraphics[width = 0.6\textwidth]{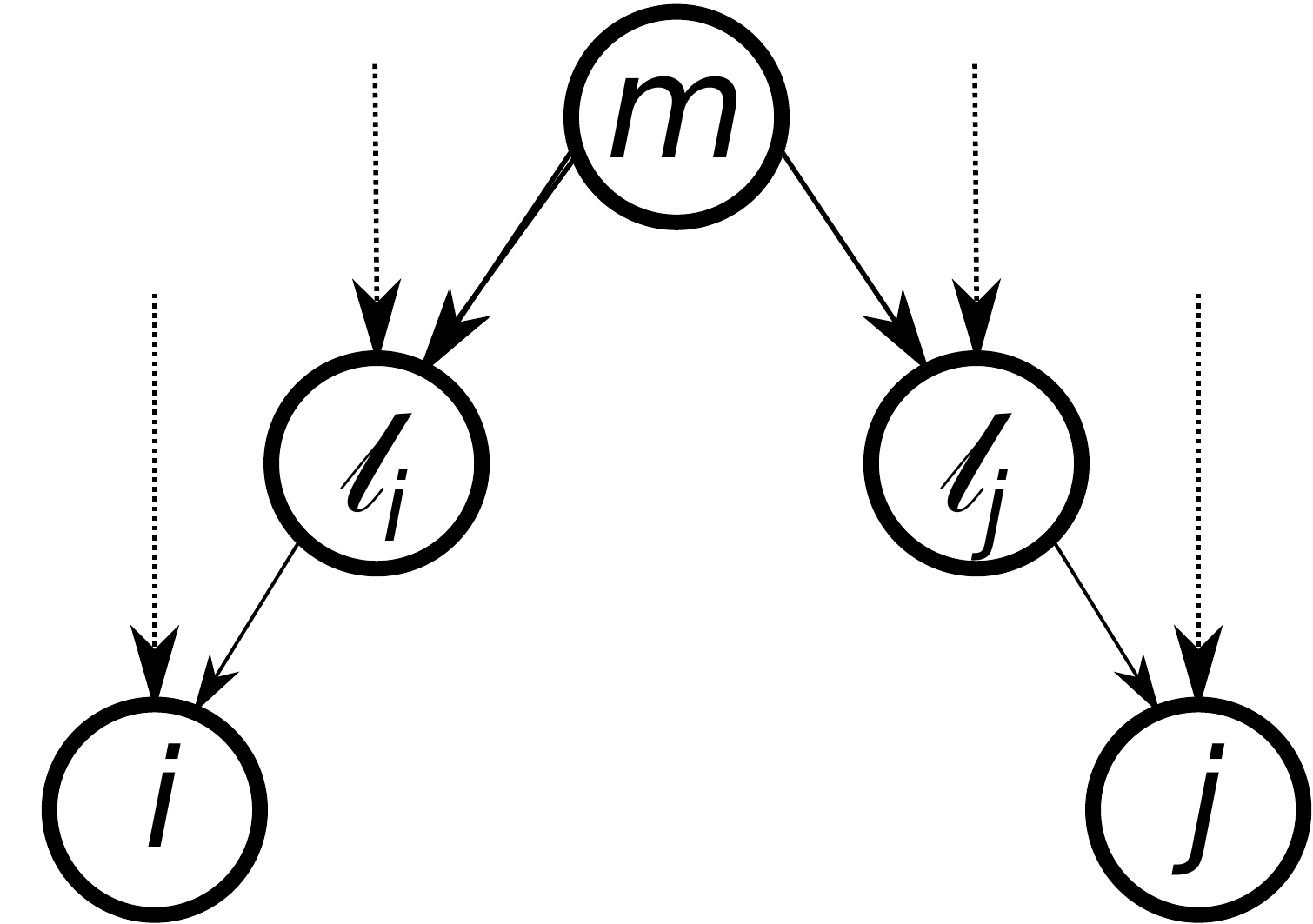}
\caption{}
\end{subfigure}
\begin{subfigure}{0.48\textwidth}
\centering
\includegraphics[width = \textwidth]{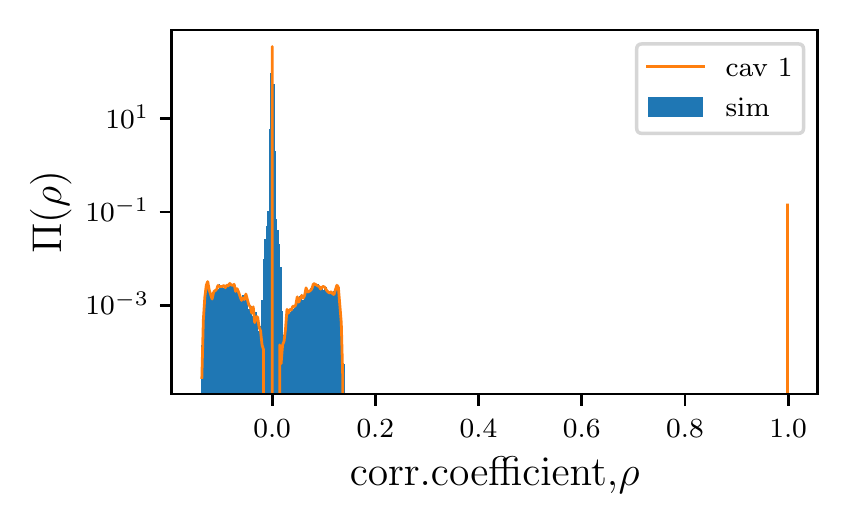}
\caption{}
\end{subfigure}
\begin{subfigure}{0.48\textwidth}
\centering
\includegraphics[width = \textwidth]{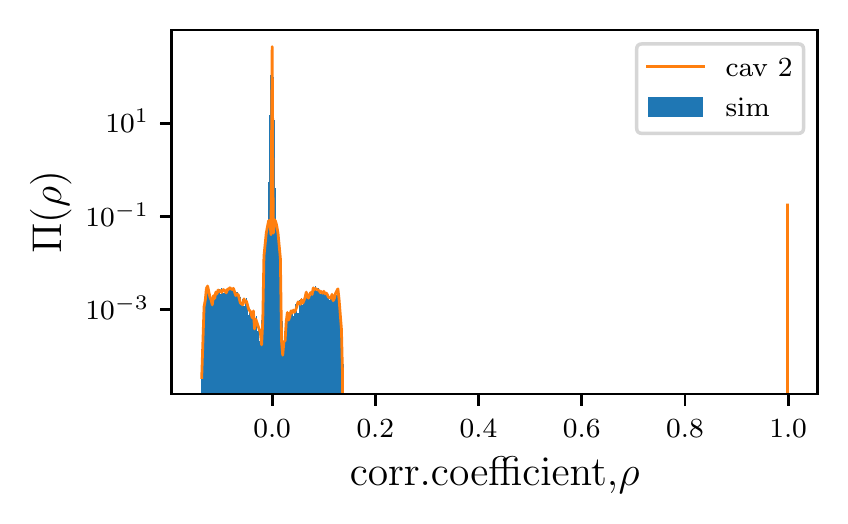}
\caption{}
\end{subfigure}
\caption{(a)-(b) Network motifs that contributes to correlation. (c)-(d) Probability density of the Pearson correlation coefficient $\rho$. Only elements of the upper triangular matrix (including the diagonal) of $\rho$ are taken into account for the histogram. Parameters: $k^{\mathrm{in}}= 2$. $N = 5000$, $T/J=0.4$, $\vartheta=0$. }
\label{fig:correlation}
\end{figure}

Let us consider the stationary (time-translation invariant) pairwise connected correlation (also called covariance)
\begin{equation}
C_{ij}^{\mathrm{c}}(\tau)\coloneqq 
\lim_{t\rightarrow\infty}\,
\langle n_i(t+\tau)n_j(t)\rangle-P_iP_j\,,
\end{equation}
that is defined between any pair of sites $\lbrace i,j\rbrace$ and time-lag $\tau$. The empirical stationary  pairwise connected  correlation (as computed from Monte-Carlo simulations) is analogously defined as 
\begin{equation}
\hat{C}_{ij}^{\mathrm{c}}(\tau)\coloneqq \frac{1}{t_s} \sum_{t=t_0}^{t_0+t_s} n_i(t+\tau)n_j(t)-P_{i,MC}P_{j,MC}\,.
\end{equation}
for $t_s\gg 1$, and $t_0$ a sufficiently long time to allow the dynamics to relax to stationarity. The covariance $C_{ij}^{\mathrm{c}}(\tau)$ depends both on  the sites and on the lag. In the following  we restrict ourselves to two levels of analysis, namely the single-site lagged auto-covariance $C_{ii}^{\mathrm{c}}(\tau)$, and the covariance at zero time lag, $C_{ij}^{\mathrm{c}}(0)$. We compare analytical results for those two types of covariance against simulations.
\paragraph{Single-site lagged auto-covariance}
For the dynamics described by \Eref{eq: model}, we first investigate the  lagged auto-covariance $\hat{C}_{ii}^{\mathrm{c}}(\tau)$. 
We will present and discuss results in term of the time-lagged local Pearson correlation defined as, $\rho_{ii}(\tau)\defeq \hat{C}_{ii}^{\mathrm{c}}(\tau)/\hat{C}_{ii}^{\mathrm{c}}(0)$, as well as its sample average $\langle \rho(\tau)\rangle=N^{-1}\sum_i\rho_{ii}(\tau)$. 
Simulation results show that $\rho_{ii}(\tau)\approx 0 $  for $\forall i, \forall\tau> 0$ (hence trivially $\langle \rho(\tau)\rangle\approx 0$), which means that the dynamics presents no retarded auto-covariance, see Fig.\,\ref{fig:time_correlation}. This is consistent with the theoretical prediction from the cavity formalism on fully asymmetric networks in the absence of loops, 
$C_{ii}^{\mathrm{c}}(\tau)=0$
$\forall i,\forall \tau>0$. Heuristically, non-zero auto-covariance is associated with feedback loops which send the information back to the starting node with a delay. However, on a  directed tree this never happens. 
\paragraph{Pairwise connected correlation}
Below we compute the analytical  pairwise  covariance at zero time lag, $C_{ij}^c(0)$, which  we refer to as $C_{ij}^{\mathrm{c}}$ for simplicity. 
Correlations between two nodes $i,j$ are expected to arise from common ancestors of the two nodes. Given the earlier result of the absence of single-site lagged auto-covariance, we expect that 
if the common ancestor $k$ is not at the same distance from $i$ and $j$, signals would need a different amount of time to travel from $k$ to $i$ and $j$, respectively, entailing that $C_{ij}^c(\tau)\neq 0$ {\em only\/}  at a non-zero $\tau$ corresponding to the difference in distances. Here we are interested in the equal-time connected correlation, $C_{ij}^c(0)$, which is thus non-zero only if the common ancestor $k$ is at the {\em same\/} (finite) distance from nodes $i$ and $j$. Therefore, network motifs that  produce non-zero values of the pairwise equal-time covariance are required to have their common root node at \textit{equal} distance from the leaf nodes; see for example the motifs that are shown in Fig.\,\ref{fig:correlation}(a)-(b).
 The (equal-time) connected correlation is obtained from the joint expectation
\begin{equation}
C_{ij}= \left\langle n_i n_j\right\rangle = \left\langle \Phi_{\beta}\left( h_i(\bm{n}_{\partial_i})-\vartheta_i\right)\Phi_{\beta}\left( h_j(\bm{n}_{\partial_j})-\vartheta_j\right) \right\rangle_{\bm{n}_{\partial_i}\cup\,\bm{n}_{\partial_j}}\,,
\label{eq:expectation_correlation}
\end{equation} 
via $C^{\mathrm{c}}_{ij}=C_{ij}-P_iP_j$, where $\bm{n}_{\partial_i}\cup\,\bm{n}_{\partial_j}$ indicates the set of states associated with the predecessors of nodes  $i$ or $j$. In the following we consider contributions to the correlations coming from the two types of motifs shown in Fig.\,\ref{fig:correlation}(a-b). They are (a) a pair of nodes sharing one predecessor, and (b) a pair of direct descendants from the nodes in motif (a). We show that those motifs provide the dominant contributions to the $C_{ij}$ distribution. If we use the standard cavity approximation 
\begin{equation}
P(\bm{n}_{\partial_i}\cup\,\bm{n}_{\partial_j},t)\approx \prod_{\ell\in\partial_i\cup\,\partial_j} P(n_\ell,t)\,,
\label{eq:correlation_cavity}
\end{equation}
which assumes  that the joint probability of node activations of the predecessors of nodes $i$ or $j$ factorises over single-node terms, we are able to take into account the effect of motif (a) (but only motif (a)). This is because the nodes belonging to the set of predecessors of both $i$ and $j$, $\lbrace  \ell,\ell \in\partial_i\cap \partial_j\rbrace$, appear only once in the product on the r.h.s. of \Eref{eq:correlation_cavity}. This then explains why $C_{ij}^{\mathrm{c}}\neq 0$ in these cases.
We evaluate \Eref{eq:expectation_correlation} in the approximation that takes motif (a) into account through dynamic programming in a similar fashion as done above in the context of the node activation probability. Let $K = |\partial_i\cup\, \partial_j|$ be the number of joint predecessors of nodes $i$ or $j$ and, to ease the notation, let us use $\lbrace 1,\dots K \rbrace$ to enumerate the nodes in this set of predecessors.
 We define the family of recursive functions $C_{ij}^{(1)}(\ell,\tilde{h}_i,\tilde{h}_j)$ such that $\ell$ represents the node we perform the average over. If node $\ell$ is active, it gives a contribution $J_{i\ell}$ and $J_{j\ell}$ to the auxiliary fields $\tilde{h}_i$ and $\tilde{h}_j$, respectively. Otherwise  the auxiliary fields do not change.  The dynamic programming evaluation for $C_{ij}$ becomes
\begin{equation}
\begin{aligned}
C_{ij}&=C^{(1)}_{ij}(1,0,0)\\
C_{ij}^{(1)}(\ell,\tilde{h}_i,\tilde{h}_j) &= P_\ell C_{ij}^{(1)}(\ell+1,\tilde{h}_i+J_{i\ell},\tilde{h}_j+J_{j\ell})+(1-P_\ell)C_{ij}^{(1)}(\ell+1,\tilde{h}_i,\tilde{h}_j)& \mbox{for } 1\leq\ell\leq K\\
C_{ij}^{(1)}(K+1,\tilde{h}_i,\tilde{h}_j) &= \Phi_{\beta}\left( \tilde{h}_i-\vartheta_i\right)\Phi_{\beta}\left( \tilde{h}_j-\vartheta_j\right) \,
\end{aligned}
\label{eq:corr_cav_1}
\end{equation}
 where $P_\ell$ is the stationary probability of node $\ell$. From the connected correlation we obtain the Pearson correlation coefficient 
\begin{equation}
 \rho_{ij}=\frac{C_{ij}^{\mathrm{c}}}{\sqrt{(P_i-P_i^2)(P_j-P_j^2)}}\,,
 \end{equation}
 and we compare the values $\rho_{ij}$ obtained through cavity approximation with that obtained from the direct evaluation of the microscopic dynamics in Fig.\,\ref{fig:correlation}(c). For the reasons explained above, we obtain non-zero values of $\rho_{ij}$ if and only if nodes $i$ and $j$ have a common predecessor $\ell$,  as in  the motif shown in Fig.\,\ref{fig:correlation}(a).  Apart from the trivial peak at 1, resulting from the diagonal terms of the distribution, the largest absolute values of $\rho$ are captured nearly perfectly by the motifs illustrated in Fig.\,\ref{fig:correlation}(a).  On the other hand, such motifs do not capture  well  the events giving rise to small correlations.
 
 In order to improve the characterisation of these events, for every pair of nodes $\{\ell_i,\ell_j\}$ that are described by the  motif (a),  we investigate the connected correlation  associated with the pair of nodes $\{i,j\}$, with $i$ and $j$ a successor of $\ell_i$ and $\ell_j$ respectively. This motif is illustrated in Fig.\,\ref{fig:correlation}(b) and  consists of a pair of nodes $\{i, j\}$ that
 has at least one common ancestor $m$ at distance two. In the following, we use the estimate of the joint expectation for $\{\ell_i,\ell_j \}$ provided by \Eref{eq:corr_cav_1} to estimate the joint expectation of the pair $\{i,j\}$.
For the two nodes denoted $\ell_i$ and $\ell_j$ in the Figure, there is a non-zero correlation, 
 $\rho_{\ell_i \ell_j}\neq 0$ as shown above, 
 hence the joint probability of the states of the predecessors of $i$ or $j$ does not factorise, and \Eref{eq:correlation_cavity} does not hold anymore. For a given pair $\lbrace i,j\rbrace$, let us denote the set of pairs of their predecessors $\{\lbrace\ell_i,\ell_j\rbrace\}$  which have a predecessor in common by $L_1= \lbrace \{\ell_i,\ell_j\} : \ell_i\in \partial_{i},\ell_j \in\partial_{j} \mbox{ and } \partial_{\ell_i}\cap\partial_{\ell_j}\neq \varnothing   \rbrace$, and let $L_0$ be set of predecessors of $i$ or $j$ which do not appear in $L_1$ and 
 are treated as independent.
 Instead of \Eref{eq:correlation_cavity}, the marginal probability of the joint set of neighbours contains  the two-points joint probability of the pairs of states  
 in the set $L_1$
 \begin{equation}
P(\bm{n}_{\partial_i}\cup\,\bm{n}_{\partial_j},t)\approx \prod_{\ell\in L_0} P(n_\ell,t)\prod_{\ell_i,\ell_j \in L_1}P(n_{\ell_i},n_{\ell_j},t)\,.
\label{eq:correlation_cavity2}
\end{equation}
For each pair of sites in $L_1$, the joint probability   $P(n_{a},n_b,t)$  is 
\begin{multline}
P(n_{a},n_b,t)=  \Big\langle\left[ n_a \Phi_\beta\left(h_a(\bm{n}_{\partial_a}))\right)+(1-n_a)(1- \Phi_\beta\left(h_a(\bm{n}_{\partial_a}))\right)\right]\\\left[ n_b \Phi_\beta\left(h_b(\bm{n}_{\partial_b})\right)+(1-n_b)(1- \Phi_\beta\left(h_b(\bm{n}_{\partial_b})\right)\right]\Big\rangle_{{\bm{n}_{\partial_a}\cup\,\bm{n}_{\partial_b}},t-1}\,.
\end{multline}
To simplify the notation, we drop the explicit time dependence in $P(n_{a},n_b,t)$; the result can be written in terms of $C_{ab},P_a,P_b$  as
\begin{multline}
P(n_a,n_b)=(1-n_a)(1-n_b)+P_a(2n_a-1)(1-n_b)+P_b(2n_b-1)(1-n_a)\\
+C_{ab}(2n_a-1)(2n_b-1)\,,
\label{eq: two-point_marginal_explicit}
\end{multline}
and $C_{ab}$ is computed from \Eref{eq:corr_cav_1}. 
 We now evaluate \Eref{eq:expectation_correlation} 
using \Eref{eq:correlation_cavity2}, which also takes into account motifs (b), in contrast to \Eref{eq:correlation_cavity} which only 
accounts for motifs (a).
 The correction applies to the evaluation of the  entries $C_{ij}$ for which a node $m$ exists which is an ancestor of both nodes $i$  and $j$  through the paths $(i, \ell_i, m)$ and $(j, \ell_j,m)$; see  motif in  Fig.\,\ref{fig:correlation} (b). Let $K_0 = |L_0|$ and $K_1=|L_1|$, and let $L_0 = \lbrace 1,\dots K_0\rbrace$  and $L_1= \lbrace\{ a_1,b_1\},\dots \{a_{K_1},b_{K_1}\} \rbrace$. \Eref{eq:corr_cav_1}  is replaced by the dynamic programming relationships
\begin{equation}
\begin{aligned}
C_{ij}&=C_{ij}^{(1)}(1,0,0)&\\
C_{ij}^{(1)}(\ell,\tilde{h}_i,\tilde{h}_j) &= P_\ell C_{ij}^{(1)}(\ell+1,\tilde{h}_i+J_{i\ell},\tilde{h}_j+J_{j\ell})+(1-P_\ell)C_{ij}^{(1)}(\ell+1,\tilde{h}_i,\tilde{h}_j)& \mbox{for } 1\leq\ell\leq K_0 \\
C_{ij}^{(1)}(K_0+1,\tilde{h}_i,\tilde{h}_j) &= C^{(2)}_{ij}(1,0,0)\\
C^{(2)}_{ij}(\ell,\tilde{h}_i,\tilde{h}_j) &= P(n_{a_\ell}=1,n_{b_\ell}=1)C_{ij}^{(2)}(\ell+1,\tilde{h}_i+J_{ia_\ell},\tilde{h}_j+J_{j b_\ell})
\\&~~+P(n_{a_\ell}=1,n_{b_\ell}=0)C_{ij}^{(2)}(\ell+1,\tilde{h}_i+J_{i a_\ell},\tilde{h}_j)
\\&~~+P(n_{a_\ell}=0,n_{b_\ell}=1)C_{ij}^{(2)}(\ell+1,\tilde{h}_i,\tilde{h}_j+J_{j b_\ell})
\\&~~+P(n_{a_\ell}=0,n_{b_\ell}=0)C_{ij}^{(2)}(\ell+1,\tilde{h}_i,\tilde{h}_j)& \mbox{for } 1\leq\ell\leq K_1\\
C^{(2)}_{ij}(K_1+1,\tilde{h}_i,\tilde{h}_j) &= \Phi_{\beta}(\tilde{h}_i-\vartheta_i)\Phi_\beta(\tilde{h}_j-\vartheta_j)\,, 
\end{aligned}
\label{eq:corr_cav_2}
\end{equation}
where the two point marginals $P(n_a,n_b)$ are computed from \Eref{eq: two-point_marginal_explicit} using \Eref{eq:corr_cav_1} to compute $C_{ab}$.
The comparison between simulations and the analytical calculation is shown in  Fig.\,\ref{fig:correlation}(d). The results indicate that the motifs in Figs.\,\ref{fig:correlation}(a) and (b)  very accurately capture the salient properties of  pairwise correlations  in this system. This finding suggests that methods for network reconstruction from experimentally measured correlations may have limited ability to characterise motifs that involve common ancestors at distance larger than two.  

\section{Extensions}
\subsection{Multi-node interaction model}
\label{factor graph}
The linear threshold model presented above in \Eref{eq: result cavity} describes a dynamic model consisting of pairwise interactions between nodes. However,  approaches based on pairwise interactions  may not always be appropriate \cite{lambiotte2019networks,battiston2020networks}. Therefore several models that explicitly  consider multi-node interactions (called high-order models) have  been proposed,  which can be implemented in terms of hypergraphs \cite{ghoshal2009random}, simplicial complexes \cite{petri2018simplicial}, and bipartite or factor graphs.

\begin{figure}
\centering
\begin{subfigure}{0.4\textwidth}
\includegraphics[width =0.8 \textwidth]{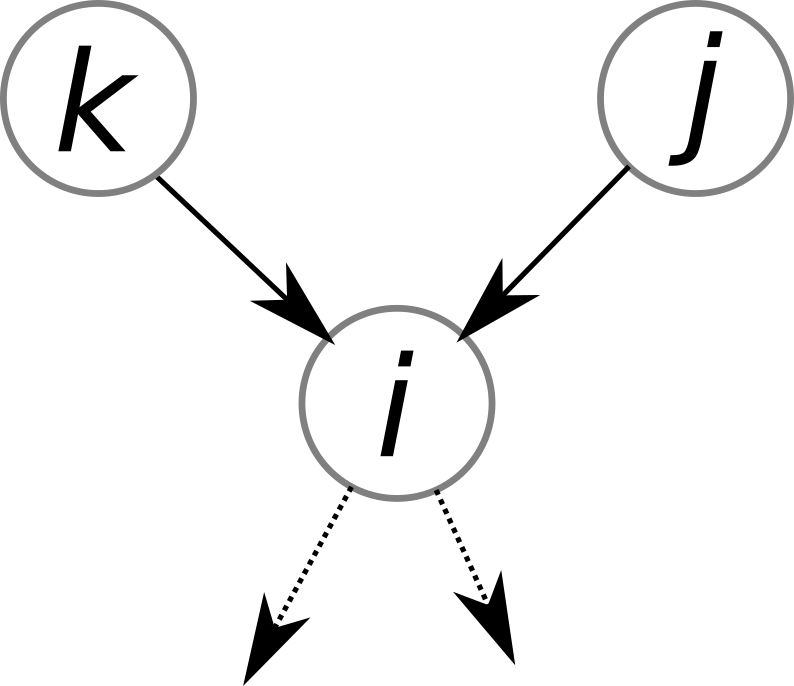}
\caption{}
\end{subfigure}
\begin{subfigure}{0.4\textwidth}
\includegraphics[width = \textwidth]{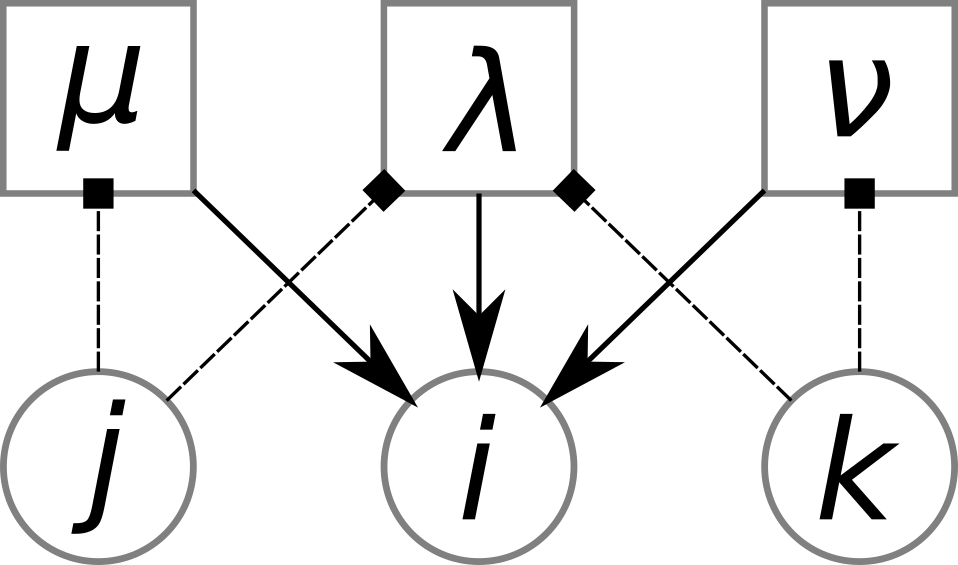}
\caption{}
\end{subfigure}
\caption{ Snapshot of a network describing  a pairwise model; each  link carries the signal from one node only (left). A bipartite graph is used to represent multi-node interaction, here  the node $i$ receives the signal from nodes $j,k$ through function nodes $\mu,\nu$ and $\lambda$, with node $\lambda$ 
representing a Boolean function of two variables (right).  Arrows indicate a link pointing from a function node to a node, square markers join a function node with the arguments of the function. }
\label{fig:bipartite_drawings}
\end{figure}

Let us consider a directed bipartite graph $G(V,E)$, where $V$ denotes the set of nodes and $E$ the set of edges. The set of nodes $V$ is composed of two disjoint sets  $X$ and $Y$, with no edges in $E$ having both endpoints in $X$ or $Y$.  Let $|X|=N$, and $|Y|=\alpha N$. We refer to the $N$ elements of $X$ as variable nodes and the $\alpha N$ elements of $Y$ as function nodes. To define a dynamic process on the bipartite network,  we associate a dynamic variable with every node of $X$. We use $\lbrace n_i \rbrace_{i=1}^N$ and $\lbrace \tau_\mu \rbrace_{\mu=1}^{\alpha N}$ to denote the dynamic variables and the function nodes respectively. Every function node $\mu$ is associated  with a Boolean function $g_\mu(\cdot)$, with the function $g_\mu(\bm{n}_{\partial _\mu})$ depending on the states of $\mu$'s predecessors $\bm{n}_{\partial _\mu}$. 
 Every variable node $i$ evolves according to a linear threshold function depending on node $i$'s predecessors $\bm{\tau}_{\partial_i}$.
The bipartite graph is defined by the bi-adjacency matrix ${\bf A}=(A_{\mu i})$ and the interaction matrix ${\bf J}=(J_{i\mu })$, with entries  independent of $\bf{A}$. The matrix ${\bf A}=(A_{\mu i})$  has dimension  $\alpha N \times N$; a value $A_{\mu i}=1$ indicates that a link from  the variable node $i$ to the function node $\mu$  exists, signifying the fact that the state $n_i$ is an argument of the function $g_\mu$, and $A_{\mu i }=0$ otherwise.  
The interaction matrix ${\bf J}=(J_{i\mu})$ has dimension $N\times \alpha N$; a non-zero entry $J_{i\mu}$  represents the strength of the influence of the function $\mu$ on the variable node $i$. 
 For any realisation of the matrices ${\bf A}$ and ${\bf J}$, we can define $d_i({\bf J})$ as the number of predecessors of node $i$ and  $c_\mu({\bf A})$  the number of predecessors of node $\mu$:
\begin{align}
d_i({\bf J})=\sum_{\mu}^{\alpha N} \Theta\left( |J_{i\mu}|\right) \,,\qquad
c_\mu({\bf A})=\sum_{i}^N A_{\mu i}\,.
\label{degreeDefs_d}
\end{align}
We consider the dynamics where the state 
of the function node $\mu$, 
given by
the binary variable $\tau_\mu$, is determined by $g_{\mu}(\bm{n}_{\partial _\mu})$. Conversely, the binary variables $n_i$ follow a linear threshold dynamics formulated in terms of the $\tau_\mu$, i.e.
 \begin{equation}
 \begin{aligned}
n_i(t+1) &= \Theta\left(\sum_{\mu}J_{i\mu}\tau_\mu(t)-\vartheta_i-z_i(t)\right)\\
\tau_\mu(t)& = g_{\mu}\left(\bm{n}_{\partial _\mu}(t)\right)\,.
\end{aligned}
\label{eq:bipartite_microscopic}
 \end{equation}
where the $z_i(t)$ represent independent identically distributed stochastic noise terms, whose  statistics are defined in terms of their cumulative distribution function $\Phi_\beta(z)$ as before.
The update rule in \Eref{eq:bipartite_microscopic} represents a more flexible model 
than
the pairwise interaction model defined in \Eref{eq: model}.  In Fig.\,\ref{fig:bipartite_drawings} (a) we show an example of a small monopartite network that captures pairwise interactions as described by \Eref{eq: model}: a node $i$ receives the signal from the two nodes $j$ and $k$ through distinct links.  Instead, the bipartite graph model shown in Fig\,\ref{fig:bipartite_drawings} (b) is more flexible, as the function nodes return arbitrary Boolean functions $g_\mu(\cdot), g_\nu(\cdot), g_\lambda(\cdot)$ of their respective inputs. These {\em can\/} represent the standard pairwise interactions through function nodes such as node $\mu$ in Fig.\,\ref{fig:bipartite_drawings}(b), which is a function of a single variable $n_j$, but also generic multi-node interactions such as through node $\lambda$, which depends on two Boolean variables $n_j$ and $n_k$.

In what follows, we develop the formalism to solve the dynamics of bipartite graphs of the type represented in Fig.\, \ref{fig:bipartite_drawings} (b), and we show that these systems  benefit from the speedup of the algorithm discussed above. We will specifically look at an example of a system with multi-node interactions  that is motivated by  gene regulations \cite{Torrisi+20}.
In the following  we investigate the case $g_\mu (\bm n) = \prod_{i\in \partial_\mu} n_i$, which represents an ``AND'' gate logic,
\begin{equation}
 \begin{aligned}
n_i(t+1) &= \Theta\left(\sum_{\mu}J_{i\mu}\tau_\mu(t)-\vartheta_i-z_i(t)\right)\\
\tau_\mu(t)& = \prod_{j \in\partial_\mu} n_j(t)\,.
\end{aligned}
\label{eq:AND_microscopic}
 \end{equation}
 
Let $P_i(t)$ be the probability that  state $n_i(t) =1$. Let $\mathcal{P}_\mu(t) \coloneqq \mathrm{Prob}\left[\tau_\mu(t) = 1\right]$ be the probability that  state $\tau_\mu(t) = 1$. If the directed bipartite graph does not contain short loops, the dynamic cavity method provides an efficient approximate solution of the dynamics. In particular the node activation probabilities are given by the set of expressions
\begin{align}
P_i(t+1)&=\left \langle  \Phi_\beta\left(h_i(\pmb{\tau}_{\partial_i})-\vartheta_i\right)\right \rangle_{\bm{\tau}_{\partial_i} }\label{eq:non_linear_prob}\\
\mathcal{P}_\mu(t) &= \prod_{j \in\partial_\mu} P_j(t)
\label{eq:non_linear_prob2}
\end{align}
with $h_i(\bm{\tau}_{\partial_i}) =\sum_\mu J_{i\mu }\tau_\mu$.
\Eref{eq:non_linear_prob} is of the same form as \Eref{eq: result cavity} and thus the dynamic programming approach can speed-up the evaluation of the average. Let us consider a node $i$ with in-degree $ d_i$, and let $\nu \in \lbrace 1,\dots d_i\rbrace$ be the indices of the predecessors of node $i$. We define a function $f_i(\nu,\tilde{h})$ recursively 
\begin{equation}
   \begin{aligned}
   f_i(\nu,\tilde{h}) &= \mathcal{P}_\nu(t) f_i(\nu+1,\tilde{h} +J_{i\nu})+(1-\mathcal{P}_\nu(t)) f_i(\nu+1,\tilde{h}) &   \mbox{for }  1\leq \nu\leq d_i\\
   f_i(d_i+1,\tilde{h}) &= \Phi_{\beta} \left(\tilde{h} -\vartheta\right)\ .
   \end{aligned}
   \label{eq: recursive_factor_graph}
   \end{equation}
In terms of this recursion, we have
\begin{equation}
    P_i(t+1) = f_i(1,0)\ .
\end{equation}
 In analogy with the pairwise model, we assume that  $J_{i\mu}\in \lbrace-J_i,0,J_i \rbrace$ with $J_i$ a real positive number that may depend on the variable node. Under this condition, the dynamic programming algorithm provides a speedup  of the evaluation of node activation probabilities.
 
\begin{figure}
\includegraphics[width = \textwidth]{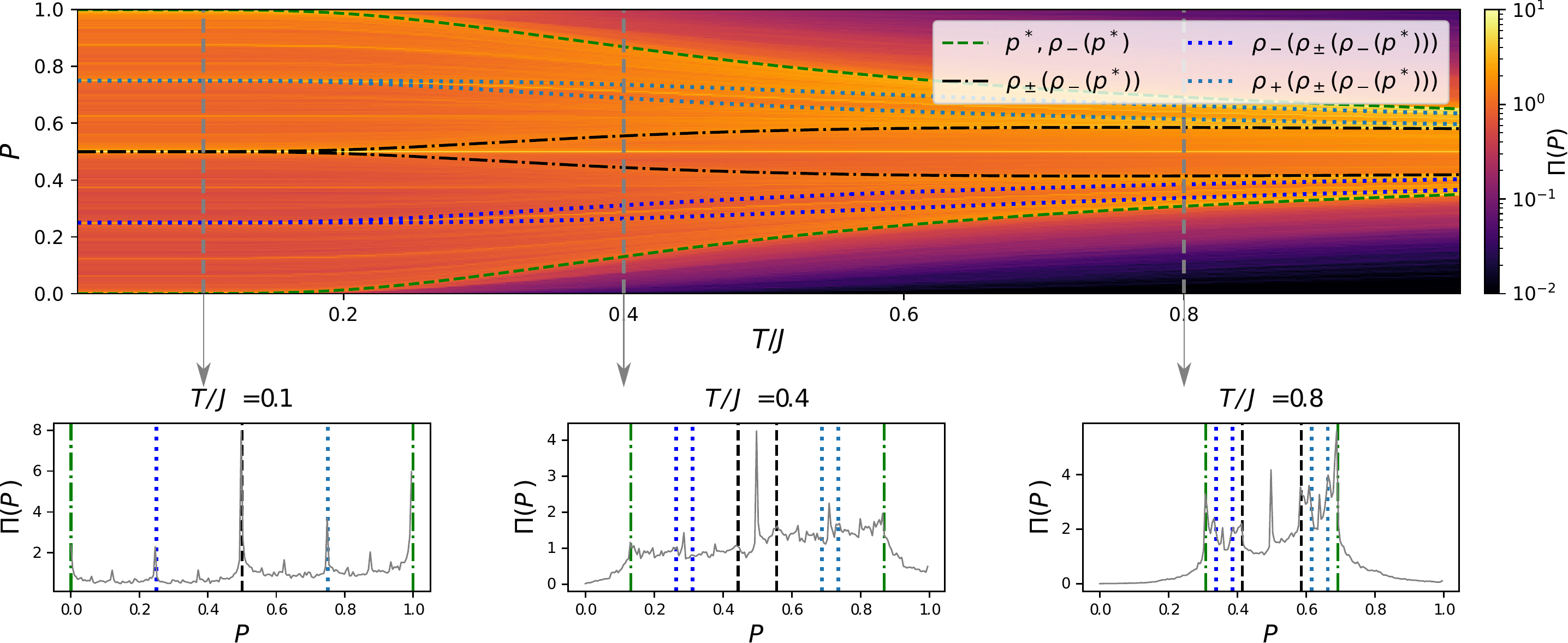}
\caption{ Heat-map   of the distribution of  node activation probabilities  $\Pi(P)$  at different values of noise parameter $T$ resulting from the microscopic dynamics detailed in \Eref{eq:AND_microscopic} (top). 
Dashed, dotted,  and dot-dashed  lines show the graph of different function compositions of  $\rho_\pm$ at different values of $T$, see \Eref{eq:analytical_map}. Vertical dotted lines mark the noise levels at which the histograms of $\Pi(P)$ are shown in the bottom  panels.  They   probe the regimes at low, intermediate, and high noise corresponding to  $T/J=0.1, 0.4, 0.8$,  from left to right. Parameters: 
$N = 200000$, $\alpha =1$, $\gamma = 2.81$, $\vartheta = 0$. }
\label{fig:heatmap_AND}
\end{figure}

 \paragraph{Node activation probability with multi-node interaction} 
 As for the linear threshold model, we inspect the stationary state of the node activation probabilities at different values of the noise intensity; results are shown in  Fig.\,\ref{fig:heatmap_AND}. In our investigation, we use the same degree distribution of the in- and out-degree as in the monopartite network, both for the variable and function nodes. The variable $\eta$ refers to the fraction of positive terms in the interaction matrix $\bm{J}$ in analogy to the monopartite case. 

Dominant peaks of the node activation probability can be rationalised in terms of the discrete stochastic map of \Eref{eq:analytical_map}. This should not come as a surprise: given our choice of the degree distribution, the majority of nodes only have a pairwise interaction with a single predecessor node. Therefore the same argument used to derive  the theoretical lines of \Eref{eq:analytical_map}, which describes the temperature dependence of some of the prominent peaks in the distribution of node activation probabilities, still holds. However, the distribution shown in Fig.\,\ref{fig:heatmap_AND} is substantially different from the distribution obtained in the linear threshold model at zero thresholds,  which was analysed  in Ref.\,\cite{torrisi2021overcoming}, highlighting contributions due to multi-node interactions. The effect of multi-node interactions  is particularly easy to analyse in the high noise regime, which is discussed in  \ref{sec:high noise}.

 \subsection{Any Boolean function can be written in terms of AND gates}
\label{sec:mapping_to_AND}
We have demonstrated above, how our linear programming method can be extended to analyse the stochastic dynamics of multi-node interaction models, using a formalisation in terms of bipartite graphs in which all function nodes are logical AND gates, i.e., we have $\tau_\mu = g_\mu(\bm{n}_{\partial_\mu})= \prod_{j\in \partial_\mu}n_j$ for all $\mu$. We now show that the bipartite model with AND gates is {\em universal\/} in the sense that a local field of the form $h_i = \sum_\mu J_{i\mu} \tau_\mu$ as it appears in \Eref{eq:AND_microscopic} can represent {\em arbitrary\/} functions of $k_i =|\partial_i|$ Boolean variables. To see this, consider an arbitrary real function $g:\lbrace 0,1\rbrace^k\rightarrow \mathbb{R}$ of $k$ Boolean variables. It can always be written using an expansion of the form
\begin{equation}
 g(\bm{n}) = a^{(0)}+\sum_i^k a^{(1)}_i n_i+\sum_{i<j}^{k}a^{(2)}_{ij}n_{i}n_j+\sum_{i<j<\ell}^{k}a^{(3)}_{ij\ell}n_{i}n_jn_\ell+\dots+a^{(k)}\prod_{\ell=1}^kn_\ell\ ,
 \label{eq:g_expansion}
\end{equation}
in which terms are arranged by increasing order, with suitable coefficients $a^{0}, \bm{a}^{(1)}, \bm{a}^{(2)}, \dots, a^{(k)}$, which can be found recursively  by considering values of $g(\bm{n})$  successively for  configurations  $\bm n$, with non-zero components on subsets of $\{1,\dots, k\}$ of increasing size.
The local field $h_i$ in \Eref{eq:AND_microscopic} is exactly of this form if the $\mu \in \partial_i$, i.e. the $\mu$ for which $J_{i\mu} \neq 0$, are arranged in order of increasing $|\partial_\mu|$.

\subsection{Spin dynamics }
Let $\sigma_i$, $i \in \lbrace 1,\dots N \rbrace$, denote a set of Ising spin variables, taking values in $\lbrace -1,1\rbrace$, located on the $N$ nodes of a complex network,  and let $J_{ij}$ denote the  strength of the interaction from node $j$ to node $i$.
The Ising spin dynamics is defined as 
\begin{equation}
    \sigma_i(t+1)=\mathrm{sgn} \left[\sum_{j}^{ N} J_{ij} \sigma_j(t) -\vartheta_i-z_i(t)\right],
    \label{eq: spinmodel}
\end{equation}
where $\vartheta_i$ is a threshold, and $z_i(t)$ is an independent identically distributed random variable defined as above with $
\mathrm{Prob}[z<x]=\Phi_\beta(x)$.\footnote{Note that, in analogy with the rest of the manuscript, the variables $\vartheta_i$ represent a threshold and not an external field.} 
The average of $\mathrm{sgn}(x-z)$ over the noise distribution can be expressed in terms of the cumulative distribution function of the noise as 
\begin{equation}
   \tilde{\Phi}_\beta(x)\defeq\langle \mathrm{sgn} (x-z)\rangle_z =2\Phi_\beta(x)-1\,.
\end{equation}
Let us denote by $m_i(t) \defeq \sum_{\sigma}P_i(\sigma,t)\sigma =\langle \sigma_i  \rangle_{\sigma_i,t}$ the time dependent average of the Ising spin at node $i$. For a fully asymmetric graph, the  cavity method provides an expression for the magnetisation in a similar form as above. Performing the average in 
\eqref{eq: spinmodel} gives
\begin{equation} m_i(t+1)=
\left\langle \tilde{\Phi}_\beta\left(h_i(\bm{\sigma}_{\partial_i})-\vartheta_i \right)\right\rangle_{\bm{\sigma}_{\partial_i},\,t}\ ,
\label{eq: result cavity spin}
\end{equation}
in which  $h_i(\bm{\sigma}_{\partial_i})=\sum_jJ_{ij}\sigma_j$ and $\bm{\sigma}_{\partial_i}$  plays a role analogous to \Eref{eq:average_definition} for spin variables. In analogy to the linear threshold model, \Eref{eq: spinmodel} can be evaluated efficiently through the recursion
\begin{equation}
   \begin{aligned}
   m_i(t+1) &= f_i(1,0)\\
   f_i(\ell,\tilde{h}) &= \dfrac{1+m_\ell(t)}{2} f_i(\ell+1,\tilde{h} +J_{i\ell})+\dfrac{1-m_\ell(t)}{2} f_i(\ell+1,\tilde{h}-J_{i\ell}) &   \mbox{for }  1\leq \ell\leq k_i\\
   f_i(k_i+1,\tilde{h}) &= \tilde{\Phi}_\beta\left(\tilde{h} -\vartheta_i\right)\,.
   \end{aligned}
   \label{eq: recursive_factor_graph_spin}
   \end{equation}
  \paragraph{ Mixed $p$-spin model}
  We briefly specify the expression for the p-spin model, whose equilibrium properties have long been investigated in the literature \cite{KirkThir87a}.
  Let us consider a system defined on a bipartite network with state variables  $\sigma_i\in \{\pm 1\}$  and $\tau_\mu\in \{\pm 1\}$. The update rule is 
 \begin{equation}
 \begin{aligned}
\sigma_i(t+1) &= \mbox{sign}\left(\sum_{\mu}J_{i,\mu}\tau_\mu(t)+z_i(t)-\vartheta_i\right)\,\\
\tau_\mu(t)& = \prod_{j \in\partial_\mu}\sigma_j(t)\,.
\end{aligned}
\label{eq:psin_microscopic}
 \end{equation}
Let $m_\mu(t) = \left\langle  \tau_{\mu}\right\rangle_{\tau_{\mu},t}$, and $m_i(t) = \left\langle  \sigma_{i}\right\rangle_{\sigma_{i},t}$ denote the time dependent averages of the $\tau_\mu$ and the $\sigma_i$, respectively.  The time evolution of the local magnetizations $m_i(t)$ is given by
\begin{equation}
\begin{aligned}
m_i(t+1)&=\left \langle  \tilde\Phi_\beta\left(h_i(\pmb{\tau}_{\partial_i})-\vartheta_i\right)\right \rangle_{\bm{ \tau}_{\partial_i},t}\,,\\
m_\mu(t) &= \prod_{j \in\partial_\mu} m_j(t)\,,
\end{aligned}
\label{eq:pspin}
\end{equation}
with $h_i(\bm{\tau}_{\partial_i}) =\sum_{\mu\in {\partial_i}} J_{i\mu }\tau_\mu$.  \Eref{eq:pspin} can be efficiently evaluated  using the recursive procedure below
\begin{equation}
\begin{aligned}
m_i(t+1) &=  f_i(1,0)\\
 m_i(\nu,\tilde{h}) &= \frac{1+m_\nu(t)}{2} f_i(\nu+1,\tilde{h} +J_{i\nu})+\frac{1-m_\nu(t)}{2} f_i(\nu+1,\tilde{h}- J_{i\nu})& \mbox{for } 1\leq \nu\leq d_i\\
  m_i(d_i+1,\tilde{h}) &=\tilde \Phi_{\beta} \left(\tilde{h} -\vartheta_i\right)
\end{aligned}
\end{equation}

\section{Beyond fully asymmetric  networks}
Feedback loops are associated  with   multi-stability and oscillatory behaviour of dynamical systems \cite{thomas1981relation,plahte1995feedback}. These two phenomena are often observed  also in   biological systems. Indeed, feedback mechanisms  represent an essential ingredient of biological networks, \textit{e.g.}, in  the context of gene regulation \cite{freeman2000feedback,zhang2012noise,chakravarty2021systematic}. However,  it is not yet clear what happens in the case  where  a combination of  many of these patterns is present in a network. 
 The presence of  bi-directional links with opposite signs of the interactions (such as in predator-prey systems) have been shown to produce oscillatory behaviour in the context of continuous  linear dynamics \cite{mambuca2020dynamical}, but little is known in the context of discrete state dynamics, e.g., for the linear threshold model. The  dynamics investigated in Sect.\,\ref{sec:node_heterogeneity} is defined on tree-like fully asymmetric network models, where feedback loops are absent.   

In this section, we consider the  linear threshold model  of \Eref{eq: model} in the presence of bi-directional links.  Any bi-directional link can be considered as a de-facto short loop of length two of the network. As expected,  the presence of these short loops  makes the use of the dynamic cavity method more challenging. In the cavity graph, where a node and its links have been removed, the presence of a short loop causes a  retarded self-interactions that  make the dynamics  of the cavity variables non-Markovian; see Ref.\,\cite{neri2009cavity,mimura2009parallel} and \ref{sec:derivation OTA}. An approximation scheme, called the one-time approximation, has been proposed with the intent to reduce the computational complexity created by this form of non-Markovianity \cite{neri2009cavity,aurell2011message}. The approximation  factorises the memory kernel to include a one-time-step memory term. It exactly describes the full dynamics of fully asymmetric networks and the equilibrium steady-state of systems with symmetric interactions in the replica symmetric phase. Furthermore, the one-time approximation has been  shown to effectively capture  the stationary state also of systems with partially symmetric links in the high noise regime \cite{aurell2011message,del2015dynamic}. However, the literature has so far focused on the agreement of macroscopic observables, such as the mean activation, while the distribution of individual node activations has generally been overlooked.  In this section, we investigate the heterogeneous node activation in networks  with bi-directional links using the OTA and compare the results with simulations. Among other things, our investigation also leads to an unexpected observation: the stationary state of networks with symmetric and anti-symmetric interactions are biased towards the active and the inactive states, respectively.

\subsection{The one-time-step approximation }
In this section we will only provide brief account of the OTA, leaving a more extensive discussion to the \ref{sec:derivation OTA}. Let us consider the model  \Eref{eq: model} in the case where the network of interactions contains  bi-directional links. For any node $i$ in the graph, the probability of the state at time $t+1$ depends on the states of node $i$  neighbours, $\bm{n}_{\partial_i}$, at the previous time-step $t$ through the transition probability $W[n_i|h_i(\bm{n}_{\partial_i}),\vartheta_i]$, see \Eref{eq: dynamical magnetisation}. If links are bi-directional, the states of nodes  $\bm{n}_{\partial_i}$ at time $t$ are not independent, because the state of every node $j \in \partial_i$ depends on  the past history states of node $i$ through the terms $J_{ji}n_i(s)$ for $s \in \{t-1, t-3\dots \}$.\footnote{Bi-directional links are the only origin of loops present on a tree, thus every node depends on its own history at even time lag.} In order to describe the dynamics of the nodes $j\in \partial_i$ on the cavity graph from  which node $i$ and connections to it are removed, it is useful to introduce the time-dependent cavity thresholds $\vartheta_j^{(i)}(s)\defeq \vartheta_j-J_{ji}n_i(s)$  for $s = 0,\dots t-1$ and $\vartheta_j^{(i)}(0)\defeq \vartheta_j$, which encodes the trajectory of  node $i$ in terms of threshold terms, and the  local field in the cavity graph \begin{equation}
h_j^{(i)}\left(\bm{n}^{s}_{\partial_j}\right)\defeq  \sum_{\ell\in \partial_j \setminus {\lbrace i \rbrace}}J_{j\ell}n_\ell^s\,,
\end{equation}
where here and in what follows we have denoted $n_j^s=n_j(s)$ to make the notation more compact.
Let us call  $\bm{n}_j^{0,\dots, t}$ the trajectory of node $j$ from time $0$ to $t$. For every bi-directional link $(i,j)$, the dynamic cavity method provides a recursive expression for the conditional probability 
$P_j^{(i)}\left(\bm{n}_j^{0,\dots,t}|\bm{\vartheta}^{(i)0,\dots,t-1}_j\right)$
of the trajectory of node $j$   
in terms of  the time-dependent thresholds $\bm{\vartheta}^{(i)0,\dots,t}_j=(\vartheta_j^{(i)}(0),\dots,\vartheta_j^{(i)}(t))$,   as discussed in  \cite{neri2009cavity,mimura2009parallel} and  in \ref{sec:derivation OTA}.
The OTA assumes that, for every  $j\in \partial_i$, the probability of the trajectory factorises into one-time conditional probabilities  
\begin{equation}
P_j^{(i)}\left(\bm{n}_j^{0,\dots,t}|\bm{\vartheta}^{(i), 0,\dots,t-1}_j\right)  \approx P^{(i)}_j(n_j^0)\prod_{s=0}^{t-1} P^{(i)}_j(n_j^{s+1}|\vartheta^{(i),s}_j)\,.
\label{eq:Ota factorisation}
\end{equation}

The factorisation procedure proposed in  \Eref{eq:Ota factorisation}  is not sufficient to fully specify  the cavity and marginal probability, but an additional closure condition is required,  as shown in \ref{sec:derivation OTA}.  In \ref{sec:derivation OTA} we re-investigate the derivation of the OTA, we show that different choices of the closure conditions lead to the different versions of the OTA known in the literature \cite{zhang2012inference,aurell2012dynamic}, and we also propose a new version of the  closure condition.   It has been shown that, for symmetric couplings, the version of Ref.\,\cite{aurell2012dynamic} admits the equilibrium state as a solution, while the version of Ref.\,\cite{zhang2012inference} has not been designed with that goal. We compare the performance of OTA under different closure conditions, including the accuracy in characterising the individual node activation probabilities, as well as the computational complexity of their implementation. We show  the version of Ref.\,\cite{aurell2012dynamic} outperforms  the version of Ref.\,\cite{zhang2012inference}  at equilibrium, but the situation is reversed in the non-equilibrium model we investigate. However, our results indicates the version of Ref.\,\cite{zhang2012inference} is more consistent through the different network symmetries investigated. Based on our assessment, we expect  the version of Ref.\,\cite{zhang2012inference} to provide the best trade-off between accuracy and ease of implementation  to tackle the non-equilibrium stationary state.  The closed-form expressions for the single-time marginal probability of  the OTA we will use are \footnote{In the following, we drop the time dependence from the cavity thresholds as this simplifies notation and does not lead to any ambiguity.}
\begin{equation}
P_i\left( n_i^t|\vartheta_i\right)= \sum_{n_i^{t-2}}\sum_{\bm{n}^{t-1}_{\partial_i}}W\left[n_i^t|h_i\left(\bm{n}_{\partial_i}^{t-1}\right),\vartheta_i\right]\left[\prod_{j\in \partial_i} P_j^{(i)}\left(n_j^{ t-1}|\vartheta_j-J_{ji}n_i^{t-2}\right)\right]P_i(n_i^{t-2}|\vartheta_i)\,,
\label{eq:cav_gen2}
\end{equation}
and for the cavity marginal,
\begin{equation}
P_j^{(i)}\left( n_j^{t-1}|\vartheta_j^{(i)}\right)= \sum_{n_j^{t-3}}\sum_{\bm{n}^{t-2}_{\partial_j\setminus i}}W\left[n_j^{t-1}|h_j^{(i)}\left(\bm{n}^{t-2}_{\partial_j}\right),\vartheta_j^{(i)}\right]\left[\prod_{\ell\in \partial_j\setminus i} P_\ell^{(j)}\left(n_\ell^{ t-2}|\vartheta_\ell-J_{\ell j}n_j^{t-3}\right)\right]P_j(n_j^{t-3}|\vartheta_j)\,.
\label{eq:cav_gen}
\end{equation}
The possible values of $\vartheta_j^{(i)}$   that appear in \Eref{eq:cav_gen} are  $\vartheta_j^{(i)} = \lbrace \vartheta_j, \vartheta_j -J_{ji }\rbrace$. To solve the dynamics over the trajectory of length  $t$, the initial conditions for the marginal and cavity probabilities  $P_j(n_j^0|\vartheta_j)$, $P_j^{(i)}(n_j^0|\vartheta_j)$, $P_j^{(i)}(n_j^0|\vartheta_j-J_{ji})$ are set for $\forall j$ and $\forall i : J_{ij}\neq 0$, and $P_j(n_j^1|\vartheta_j)$ is computed from \Eref{eq:cavity_0}. To evaluate the trajectory of node activation, \Eref{eq:cav_gen} and \Eref{eq:cav_gen2} are propagated over the time-steps of interests $0,\dots t$. At every time iteration, both  \Eref{eq:cav_gen} and \Eref{eq:cav_gen2} would face a complexity barrier created by a large in-degree of node $i$ and $j$ respectively, and their evaluation  benefit from our dynamic programming implementation. 
To this end, we note that the term $\sum_{\bm{n}^{t-1}_{\partial_i}} W[\ldots][\prod_{j\in \partial_i}\ldots]$ in \Eref{eq:cav_gen2} has the same structure as \Eref{eq:average_definition} and can be evaluated using dynamic programming as explained in Sec.\,\ref{sect:comparison}. This also applies to the corresponding average 
in \Eref{eq:cav_gen}.

In this section, we apply the OTA to networks with fat-tailed degree distributions and bi-directional links. We will investigate three different degrees of symmetry of the interaction matrix, namely antisymmetric, symmetric and uncorrelated interaction matrix $J$, and we highlight how  the symmetry affects the performance of the OTA in the regime of low temperature. 
\subsection{Comparison between OTA and  simulations} 
\begin{figure}
\includegraphics[width = \textwidth]{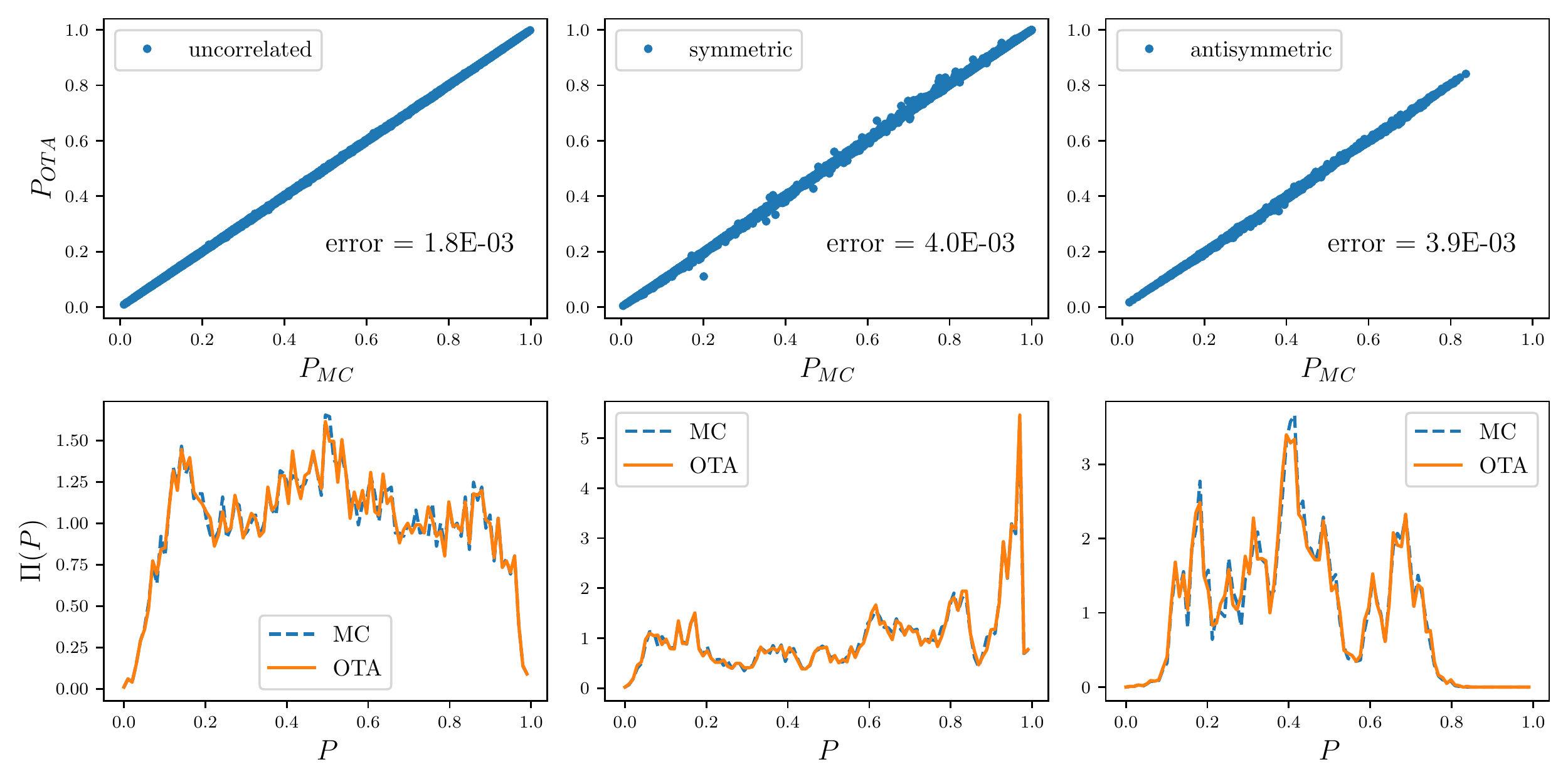}
\caption{Comparison of the one-time approximation (OTA) and Monte Carlo (MC) simulations at stationarity in networks with bi-directional links and weights chosen, from left to right, as uncorrelated, symmetric and anti-symmetric. Top panels show scatter plots of single-node activation probabilities computed from OTA versus MC results. Bottom panels show the distribution of node activation probabilities obtained through MC and OTA. Simulations are obtained through an average of over 500 trajectories. Parameters are $N = 10000$, $\gamma=4$, $T/J = 0.5,\vartheta=0$, $t_s = 10^4$ steps.}
\label{fig:comparison}
\end{figure}

\paragraph{Parameters of the model}
We first generate an undirected network in the configuration model class consisting of $N$ nodes. In analogy with above, degrees of the nodes are sampled from the fat-tailed distribution $P_\gamma(k)= k^{-\gamma}-(k+1)^{-\gamma}$. We then assign  interaction terms to the two directions of  bi-directional links.   In particular,  interaction terms may or may not have opposite sign in the two directions, \textit{i.e.} $J_{ij}J_{ji}=\pm J^2$. In the following, we investigate the correlation of the sign in bi-directional links  and we focus on three major symmetries: symmetric  ($J_{ij}=J_{ji}$), uncorrelated ($\langle J_{ij}J_{ji}\rangle=0$), and antisymmetric  ($J_{ij}=-J_{ji}$) interaction matrix $J$. In all settings, the number of positive and negative interactions are statistically the same.
We iterate the OTA equations until convergence using the same criterion we adopt on  fully asymmetric networks, as  we have discussed in   Sect.\,\ref{sect:comparison}.
\paragraph{Results}
 In order to compare the performance of OTA against Monte Carlo simulation, we define $P_{MC}$ and $P_{OTA}$ the probability of node activation derived from microscopic Monte Carlo simulation of the dynamics and from Eqs.\,\eqref{eq:cav_gen2}\eqref{eq:cav_gen}  respectively at stationarity. 
In Fig.\,\ref{fig:comparison} we compare $P_{MC}$ and $P_{OTA}$ for bi-directional interaction matrices with the three different types of symmetries detailed above. We recall that in the symmetric setting,  the stationary state of the dynamics is an equilibrium state, and in the absence of replica symmetry breaking the equilibrium state is described by belief propagation \cite{mezard2001bethe}.   
To evaluate  the accuracy of the OTA, we consider  the mean square difference of  $P_{OTA}$ to $P_{MC}$
\begin{equation}
\mathrm{error}=\sqrt{\frac{1}{N}\sum_i\left( P_{i,\mathrm{MC}}-P_{i,OTA}\right)^2}\,.
\label{eq: error generic}
\end{equation}
Our results suggest that the OTA is able to capture the statistics of node activation in the stationary state remarkably well, even in the regime of moderately low temperatures when bi-directional links are present, both for the symmetric and the uncorrelated case. However, at still lower temperature we expect the results from OTA to be less accurate, as observed in earlier work for spin models \cite{aurell2011message,aurell2012dynamic}.

\begin{figure}
\centering
\includegraphics[width =0.49\textwidth]{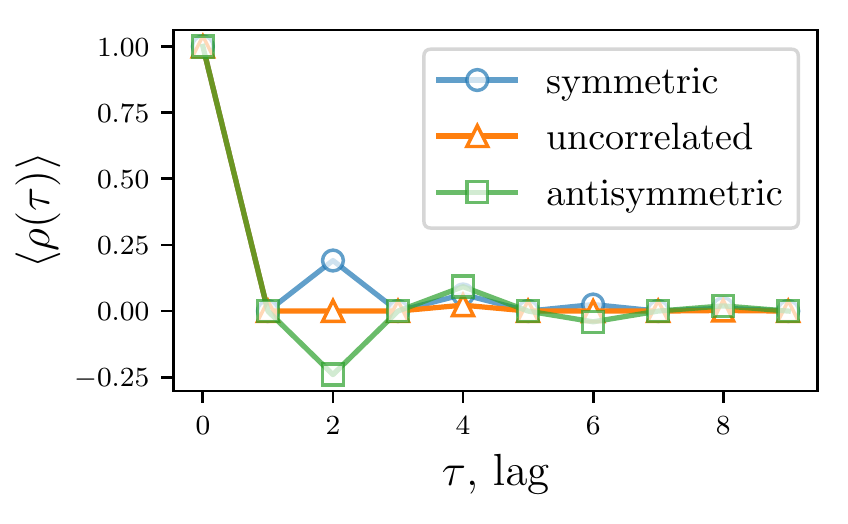}
\includegraphics[width =0.49\textwidth]{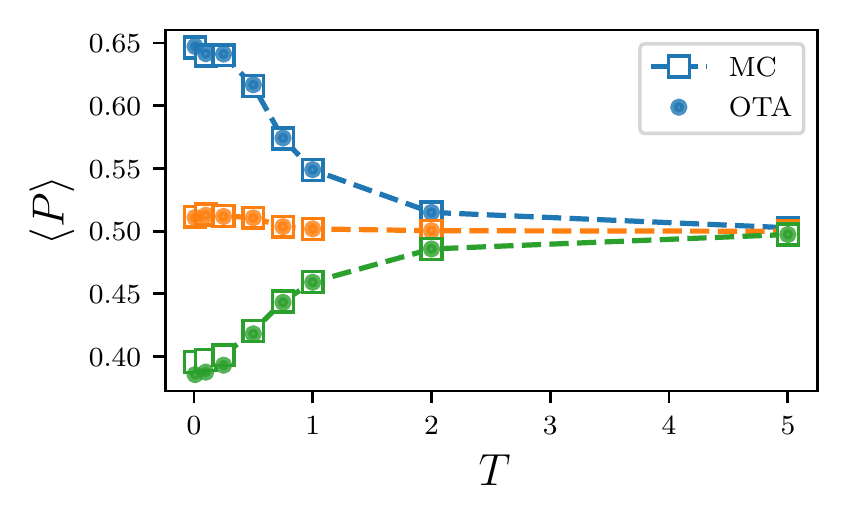}
\caption{Left: Site average of the lagged auto-correlation $\langle\rho(\tau) \rangle$ for  a network with bi-directional links. Three choices of the interaction matrix $J$ are considered: symmetric (circle), uncorrelated (triangles), and antisymmetric (squares) interaction matrices. The correlation is computed from Monte Carlo simulated trajectories of the dynamics. 
Right: site average of the node activation at different temperatures $\langle P \rangle = N^{-1}\sum_iP_i$ for symmetric (blue, upper curve), uncorrelated (orange, middle curve), and antisymmetric (green, lower curve) interaction matrices. Squares indicate the result obtained through Monte Carlo simulations, circles are obtained through the one-time approximation. Network and simulation parameters are the same as for Fig.\, \ref{fig:comparison}.}
\label{fig:correlation_bidirectional}
\end{figure}

Our earlier analysis on fully-asymmetric networks revealed that spatial correlations are short-ranged and auto-covariance is zero at any finite time lag.
On the other hand, the presence of 
 correlations in the weights   is associated with time-lagged auto-correlation at non-zero lag, as shown  in Fig.\,\ref{fig:correlation_bidirectional}.
Simulation results indicate that the antisymmetric and symmetric settings exhibit the slowest decrease of the time-lagged auto-covariance, while the uncorrelated setting exhibits the fastest. Nevertheless, the auto-covariance decays quite consistently in few steps even for the symmetric and antisymmetric interaction  models, justifying the use of OTA  to investigate the stationary state. This observation suggests that a different factorisation that includes more time-steps than in \Eref{eq:Ota factorisation} may provide a better characterisation of the symmetric and antisymmetric cases.  

Finally, Fig.\,\ref{fig:comparison} also reveals that the stationary distribution of node activations for symmetric and antisymmetric interaction is skewed, even though we are in the regime with zero thresholds, $\vartheta_i = 0$ $\forall i$, and unbiased distribution of couplings, i.e., $\langle J_{ij}\rangle =0$. We refer to this setting as ``unbiased parameter choice'' in the following.  This behaviour is confirmed in  Fig.\,\ref{fig:correlation_bidirectional}, where mean node activation are shown at different levels of the noise. In the following, we discuss this unexpected result and we provide an intuitive explanation. 
\section{Spontaneous symmetry breaking of  the linear threshold model  for unbiased parameter choices}
We have observed that, for an unbiased parameter choice, the  linear model defined in terms of $\lbrace 0,1\rbrace$ variables exhibits a skewed distribution of node activations in the presence of bi-directional links. This is somewhat surprising, as in spin systems, the average magnetisation is well known to be zero in the case of an u updatenbiased parameter choice.  In this section we compare the two models and we provide a simple example that illustrates how such a  phenomenology might.  

The dynamics described in terms of indicator variables $n_i\in \lbrace 0,1\rbrace$ can be reformulated in terms of spin variables $\sigma_i \in \lbrace \pm 1\rbrace$. 
Recall that the dynamics of the linear threshold model of \Eref{eq: model}, defined in terms  of $N$ indicator variables, 
is given by
\begin{equation}
n_i(t+1) = \Theta\left( \sum_j J_{ij}n_j(t) -\vartheta_i -z_i(t)\right)\,.
\end{equation}
Rewriting the analogous dynamics of a model of $N$ interacting spins as defined in \Eref{eq: spinmodel} as
\begin{equation}
\sigma_i(t+1) = \mathrm{sgn}\left( \sum_j J'_{ij}\sigma_j(t)- \vartheta_i' -z_i(t)\right)\, ,
\end{equation}
with interaction matrix $J'$ and thresholds $\lbrace \vartheta'_i\rbrace_{i=1}^N$,
one can map one model onto the other by setting
\begin{equation}
J'_{ij}=J_{ij}/2,\qquad \vartheta_i' = \vartheta_i-\frac{1}{2}\sum_{\ell}J_{i\ell}, \quad \forall i,j\,.
\label{eq:mapping_spin_indicator}
\end{equation}
Unless the mapping of \Eref{eq:mapping_spin_indicator} is implemented exactly, the trajectories of the Boolean linear threshold model and the spin model will be different. In analogy with the above, we use $m_i=\langle \sigma_i\rangle$ and $P_i= \langle n_i\rangle$ to denote the stationary averages of spins and indicator variables respectively.  The phase diagram for spin systems has been extensively studied at equilibrium, and it has been shown \cite{mezard2001bethe} that for an unbiased parameter choice 
 the mean magnetisation is zero, i.e., $N^{-1}\sum_i m_i=0$,
even at low temperatures.

In a system with indicator variables instead, both Monte Carlo simulations and an analysis using OTA show that the system exhibits spontaneous symmetry breaking $N^{-1}\sum_i P_i\neq 1/2$ when  the parameters  are unbiased, as shown in  Fig.\,\ref{fig:correlation_bidirectional}. In particular, at stationarity  the average site activation  $N^{-1}\sum_i P_i$ is larger than $1/2$  for symmetric interaction matrices and smaller than $1/2$ in the case of antisymmetric interaction matrices.

We present an intuitive argument that explains how the biased distribution of node activations  emerges in a model with an unbiased choice of parameters. To this end, let us consider a mini-network composed of just two nodes $i,j$ connected by a bi-directional link, i.e. $J_{ij}J_{ji}\neq 0$. A key mechanism in bi-directional links is the dependence of a node state on its own past, i.e. node $i$ interacts with its neighbours $\partial_i$ which in turn  interact with $i$. 
The probability of the state of node $i$ at time $t+2$ can be written in terms of the probability of the state of node $i$ at time $t$ using the Markov property of the microscopic dynamics,
\begin{equation}
\begin{aligned}
P_i(n_i^{t+2} )& = \left\langle W(n_i^{t+2}|J_{ij}n_j)\right\rangle_{n_j\sim P(n_j^{t+1})}\\
P_j(n_j^{t+1} ) &= \left\langle W(n_j^{t+1}|J_{ji}n_i)\right\rangle_{n_i\sim P(n_i^{t})}\,.
\label{eq:one_node_2_steps}
\end{aligned}
\end{equation}
Considering the cases $J_{ij},J_{ji} \in \{\pm J\}$, we note that $\langle W(1|Jn) \rangle_{n\sim P}= \rho_+(P)$ and   $\langle W(1|-Jn) \rangle_{n\sim P}= \rho_-(P)$  with $\rho_\pm(P)$ defined earlier in \Eref{eq:analytical_map}. In accordance with the  unbiased parameters choice, the external threshold is set to zero, $\vartheta_i =\vartheta_j =  0$. 
\paragraph{Symmetric network}
In a symmetric network it holds $J_{ij}=J_{ji}$.  Thus only the two following 
cases 
can be
realised:
\begin{itemize}
\item ferromagnetic coupling $J_{ij}=J_{ji}=J$, leading to   $P_i(t+2)=\rho_+(\rho_+(P_i(t)))$.
\item antiferromagneric coupling $J_{ij}=J_{ji}=-J$, leading to  $P_i(t+2)=\rho_-(\rho_-(P_i(t)))$
\end{itemize}
In the stationary state, the node activation probability is given by the  fixed point of the map $p^+ = \rho_+(p^+)$ in the case of ferromagnetic couplings, or $p^- = \rho_-(p^-)$ in the case of antiferromagnetic couplings. 
We now exploit the above observation by considering an unbiased network of isolated dimers, allowing us to use the above result obtained for individual dimers. The site average of the node activation probability gives 
\begin{equation}
\langle P\rangle =\frac{p^++p^-}{2}= \left(2-\frac{1}{2}\tanh^2(\beta J/2)\right)^{-1}>1/2\,.
\label{eq:symmetric_avg_site}
\end{equation}
While this result  is derived for a special class of networks, we expect  that for a generic  unbiased symmetric network  the site average activation is larger than $1/2$, as shown by Fig.\, \ref{fig:correlation_bidirectional} where Monte Carlo simulation shows that $\langle P\rangle > 1/2$ at stationarity. Indeed,  from \Eref{eq:symmetric_avg_site} this simple argument predicts that at $T=0$, $P = 2/3$ which is comparable with the simulation result in Fig.\, \ref{fig:correlation_bidirectional}  (see blue markers).
\paragraph{Antisymmetric network}
An antisymmetric network defined by $J_{ij}=-J_{ji}$ is always unbiased. Let $J_{ij}=J$ (then $J_{ji}=-J)$,   \Eref{eq:one_node_2_steps} gives
\begin{itemize}
\item   $P_i(t+2)=\rho_+(\rho_-(P_i(t)))$
\item  $P_j(t+2)=\rho_-(\rho_+(P_j(t)))$.
\end{itemize}
In the stationary state, the node activation of node $i$ satisfies $p_i=\rho_+(\rho_-(p_i))$, and the node activation of node $j$ satisfies $p_j=\rho_-(\rho_+(p_j))$.
Again, we consider a network of isolated dimers. The site average of the node activation probability gives 
\begin{equation}
\langle P\rangle =\frac{p_i+p_j}{2}=\left(2+\frac{1}{2}\tanh^2(\beta J/2)\right)^{-1}<1/2\,,
\label{eq:antisymmetric_avg_site}
\end{equation}
which suggests that in  an unbiased antisymmetric network, the site average activation is smaller than $1/2$ and it is confirmed in Fig.\, \ref{fig:correlation_bidirectional}. From \Eref{eq:antisymmetric_avg_site}, this simple argument predicts that at $T=0$, $ P =  2/5$ which is comparable with the result in Fig.\, \ref{fig:correlation_bidirectional} (see green markers).

The result of \Eref{eq:symmetric_avg_site}  is an example where the dynamics of Ising spin and indicator variables leads to macroscopically different stationary states. While  a spin model defined for unbiased parameter choices presents $N^{-1}\sum_i m_i=0$ at equilibrium, for indicator variable we have rationalised that in general $2\sum_{i}P_i-1\neq 0$.

\section{Summary and discussion}
In this paper we have solved  the stochastic dynamics of  Boolean linear threshold models for a class of networks characterised by fat-tailed  degree distribution. On fully asymmetric networks,  the distribution of the node activation probabilities in the stationary state  is not unimodal in general but  displays a rich structure. Salient features of the distribution   can in part be rationalised in terms of discrete stochastic maps resulting from the underlying network structure and dynamics.

We have adapted the  dynamic cavity method to study equal-time  pairwise correlations in the stationary state. For every pair of nodes, we have computed the correlation produced by common ancestors at a distance of one or two from the pair. 
We demonstrate that these two motifs capture most of the statistics of values, indicating that pairwise correlation is effectively a local property of the network.

Finally, we have investigated the individual node activation statistics in the stationary state for two additional classes of models, namely for models with multi-node interactions and for linear threshold models defined on networks with bi-directional links. For models with multi-node interactions, we have adopted a bipartite graph representation  in which  factor nodes represent ``AND'' gates, and we have shown that this construction is universal in the sense that it allows representing \textit{any} Boolean function.
We have previously used models of this type to describe the regulatory mechanism of transcription factors (TFs), and we have so far focused on the percolation problem \cite{Torrisi+20,Hannam+19} as a way to identify structural properties that gene regulatory networks need to exhibit in order to support multi-cellular life. The present manuscript provides an additional perspective that incorporates the role of noise and of inhibition, which  were outside the focus of our previous analysis. In the high noise limit, we observe that TFs having a large in-degree are more robust to spontaneous noise-induced activation, which suggests a potential use as noise filters. On the other hand, results from percolation theory suggest that the average in-degree of TF complexes should be small to allow sustaining a multiplicity of attractors. Hence, the combination of those results may hint at a trade-off between these antagonistic conditions being relevant for gene regulatory networks.

To study systems with bi-directional links, we apply an approximation procedure, called OTA, to analyse the stationary state. The OTA assumes a factorisation of the memory kernel which in some sense amounts to a Markov assumption. The reason why the OTA is found to be effective in describing the non-equilibrium stationary state may thus be very well be related to the fact that the lagged auto-correlation is found to be short-ranged. 

The  dynamics of the linear threshold model  with bi-directional links yields a somewhat surprising result:  a network with an unbiased distribution of $\pm J$ interactions sustains a biased distribution of node activation probability, both for symmetric and anti-symmetric interaction matrices. We provide a heuristic argument that helps to explain this phenomenon. Symmetric and anti-symmetric interactions are known to appear in biological networks as a mechanism for positive and negative feedback mechanisms respectively \cite{cardelli2016noise}. The present finding is  of potential interest in the context of  biological networks and may help explain sustained activation or quiescence of nodes. Also, it demonstrates once more how the intuition built about the Ising spins model may be misleading if one investigates models defined on indicator variables \cite{campajola2021equivalence}.

\section{Acknowledgements}
GT is supported by the EPSRC Centre for Doctoral Training in Cross-Disciplinary Approaches to Non-Equilibrium Systems (CANES EP/L015854/1).\\

\noindent
{\large\bf References}

\begin{thebibliography}{10}

\bibitem{EdwardsAnderson75}
S.~F. Edwards and P.~W. Anderson.
\newblock {Theory of Spin Glasses}.
\newblock {\em J. Phys. F}, 5:965--974, 1975.

\bibitem{SK75}
D.~Sherrington and S.~Kirkpatrick.
\newblock {Solvable Model of a Spin-Glass}.
\newblock {\em Phys. Rev. Lett.}, 35:1792--1796, 1975.

\bibitem{viana1985phase}
L.~Viana and A.~J Bray.
\newblock Phase diagrams for dilute spin glasses.
\newblock {\em jpc}, 18(15):3037, 1985.

\bibitem{kauffman1969metabolic}
S.~A Kauffman.
\newblock Metabolic stability and epigenesis in randomly constructed genetic
  nets.
\newblock {\em J. Theor. Biol}, 22(3):437--467, 1969.

\bibitem{derrida1986random}
B.~Derrida and Y.~Pomeau.
\newblock Random networks of automata: a simple annealed approximation.
\newblock {\em Europhys.~Lett.}, 1(2):45, 1986.

\bibitem{ParisiPNAS1990}
G.~Parisi.
\newblock A simple model for the immune network.
\newblock {\em Proc.~Natl.~Acad.~Sci., U.~S.~A}, 87(1):429--433, 1990.

\bibitem{AgliariEtAl2012}
E.~Agliari, A.~Barra, A.~Galluzzi, F.~Guerra, and F.~Moauro.
\newblock Multitasking associative networks.
\newblock {\em Phys. Rev. Lett.}, 109:268101, Dec 2012.

\bibitem{agliari2013immune}
E.~Agliari, A.~Annibale, A.~Barra, A.~C.~C Coolen, and D.~Tantari.
\newblock Immune networks: Multitasking capabilities near saturation.
\newblock {\em J. Phys. A Math. Theor.}, 46(41):415003, 2013.

\bibitem{Hopfield82}
J.~J. Hopfield.
\newblock {Neural Networks and Physical Systems with Emergent Collective
  Computational Abilities}.
\newblock {\em Proc.~Natl.~Acad.~Sci., U.~S.~A}, 79:2554--2558, 1982.
\newblock Reprinted in \cite{Anderson88}.

\bibitem{hertz2018introduction}
J.~Hertz, A.~Krogh, and R.~G. Palmer.
\newblock {\em {Introduction to the Theory of Neural Computation}}.
\newblock CRC Press, 2018.

\bibitem{Amit87a}
D.~Amit, H.~Gutfreund, and H.~Sompolinsky.
\newblock {Statistical Mechanics of Neural Networks Near Saturation}.
\newblock {\em Annals of Physics}, 173:30--67, 1987.

\bibitem{sollich2014extensive}
P.~Sollich, D.~Tantari, A.~Annibale, and A.~Barra.
\newblock Extensive parallel processing on scale-free networks.
\newblock {\em Phys.~Rev.~Lett.}, 113(23):238106, 2014.

\bibitem{ChalletZhang97}
D.~Challet and Y.~C. Zhang.
\newblock {Emergence of Cooperation and Organization in An Evolutionary Game}.
\newblock {\em Physica A}, 246:407--418, 1997.

\bibitem{CoolenMG05}
A.~C.~C. Coolen.
\newblock {\em {The Mathematical Theory of Minority Games--Statistical
  Mechanics of Interacting Agents}}.
\newblock Oxford University Press, Oxford, 2005.

\bibitem{Iori99}
G.~Iori.
\newblock {Avalanche Dynamics and Trading Friction Effects on Stock Market
  Returns}.
\newblock {\em Int. J. Mod. Phys. C}, 10:1149--1162, 1999.

\bibitem{Bornh01}
S.~Bornholdt.
\newblock {Expectation Bubbles in a Spin Model of Market Intermittency from
  Frustration Across Scales}.
\newblock {\em Int. J. Mod. Phys. C}, 12:667--674, 2001.

\bibitem{AnandKuehn07}
K.~Anand and R.~K\"uhn.
\newblock {Phase Transitions in Operational Risk}.
\newblock {\em Phys. Rev. E}, 75:016111, 2007.

\bibitem{HaKu06}
J.~P.~L. Hatchett and R.~K\"uhn.
\newblock {Effects of Economic Interactions on Credit Risk}.
\newblock {\em J.~Phys.~A}, 39:2231--2251, 2006.

\bibitem{weigt2000number}
M.~Weigt and A.~K Hartmann.
\newblock Number of guards needed by a museum: A phase transition in vertex
  covering of random graphs.
\newblock {\em Phys.~Rev.~Lett.}, 84(26):6118, 2000.

\bibitem{cocco2001statistical}
S.~Cocco and R.Monasson.
\newblock Statistical physics analysis of the computational complexity of
  solving random satisfiability problems using backtrack algorithms.
\newblock {\em Eur. Phys. J. B}, 22(4):505--531, 2001.

\bibitem{martin2001statistical}
O.~C. Martin, R.~Monasson, and R.~Zecchina.
\newblock Statistical mechanics methods and phase transitions in optimization
  problems.
\newblock {\em Theoretical computer science}, 265(1-2):3--67, 2001.

\bibitem{Franz+01}
S.~Franz, M.~Leone, F.~{Ricci-Tersenghi}, and R.~Zecchina.
\newblock {Exact Solutions for Diluted Spin Glasses and Optimization Problems}.
\newblock {\em Phys. Rev. Lett.}, 87:127209, 2001.

\bibitem{mezard2002analytic}
M.~M{\'e}zard, G.~Parisi, and R.~Zecchina.
\newblock Analytic and algorithmic solution of random satisfiability problems.
\newblock {\em Sci.}, 297(5582):812--815, 2002.

\bibitem{mezard2001bethe}
M.M{\'e}zard and G.Parisi.
\newblock The bethe lattice spin glass revisited.
\newblock {\em Eur. Phys. J. B}, 20(2):217--233, 2001.

\bibitem{coolen2016replica}
A.C.C. Coolen.
\newblock Replica methods for loopy sparse random graphs.
\newblock In {\em J. Phys. Conf. Ser.}, volume 699, page 012022, 2016.

\bibitem{Annibale+10}
A.~Annibale, A.~C.~C. Coolen, and G.~Bianconi.
\newblock {Network Resilience Against Intelligent Attacks Constrained by the
  Degree-Dependent Node Removal Cost}.
\newblock {\em J. Phys. A}, 43:395001, 2010.

\bibitem{altarelli2013optimizing}
F.~Altarelli, A.~Braunstein, L.~Dall’Asta, and R.~Zecchina.
\newblock Optimizing spread dynamics on graphs by message passing.
\newblock {\em J. Stat. Mech.: Theory Exp}, 2013(09):P09011, 2013.

\bibitem{lokhov2015dynamic}
A.~Y. Lokhov, M.~M{\'e}zard, and Zdeborov{\'a}.
\newblock Dynamic message-passing equations for models with unidirectional
  dynamics.
\newblock {\em Phys. Rev. E}, 91(1):012811, 2015.

\bibitem{PagKu15}
P.~Paga and R.~K\"uhn.
\newblock {Contagion in an Interacting Economy}.
\newblock {\em JSTAT}, 03:P03008, 2015.

\bibitem{li2021impact}
B.~Li and D.~Saad.
\newblock Impact of presymptomatic transmission on epidemic spreading in
  contact networks: A dynamic message-passing analysis.
\newblock {\em Phys. Rev. E}, 103(5):052303, 2021.

\bibitem{KarrerNewm10}
B.~Karrer and M.~E. J.Newman.
\newblock {Message Passing Approach for General Epidemic Models}.
\newblock {\em Phys. Rev. E}, 82:016101, 2010.

\bibitem{lokhov2014inferring}
A.~Y. Lokhov, M.~M{\'e}zard, H.~Ohta, and L.~Zdeborov{\'a}.
\newblock Inferring the origin of an epidemic with a dynamic message-passing
  algorithm.
\newblock {\em Phys.~Rev.~E}, 90(1):012801, 2014.

\bibitem{roudi2011dynamical}
Y.~Roudi and J.~Hertz.
\newblock Dynamical tap equations for non-equilibrium ising spin glasses.
\newblock {\em J. Stat. Mech.}, 2011(03):P03031, 2011.

\bibitem{zhang2012inference}
P.~Zhang.
\newblock Inference of kinetic {I}sing model on sparse graphs.
\newblock {\em J. Stat. Phys}, 148(3):502--512, 2012.

\bibitem{aurell2012dynamic}
E~Aurell and H.~Mahmoudi.
\newblock Dynamic mean-field and cavity methods for diluted ising systems.
\newblock {\em Phys. Rev. E}, 85(3):031119, 2012.

\bibitem{Dominicis78}
C.~De Dominicis.
\newblock {Dynamics as Substitute for Replicas in Systems with Quenched Random
  Impurities}.
\newblock {\em Phys. Rev. B}, 18:4913--4919, 1978.

\bibitem{Hatchett+04}
J.~P.~L. Hatchett, B.~Wemmenhove, I.~P\'{e}rez~Castillo, T.~Nikoletopoulos,
  N.~S. Skantzos, and A.~C.~C. Coolen.
\newblock {Parallel Dynamics of Disordered Ising Spin Systems on Finitely
  Connected Random Graphs}.
\newblock {\em J.~Phys.~A}, 37:6201--6220, 2004.

\bibitem{mimura2009parallel}
K.~Mimura and A.~C.~C. Coolen.
\newblock Parallel dynamics of disordered ising spin systems on finitely
  connected directed random graphs with arbitrary degree distributions.
\newblock {\em J. Phys. A}, 42(41):415001, 2009.

\bibitem{neri2009cavity}
I.~Neri and D.~Boll{\'e}.
\newblock The cavity approach to parallel dynamics of ising spins on a graph.
\newblock {\em J. Stat. Mech.: Theory Exp}, 2009(08):P08009, 2009.

\bibitem{KuRog17}
R.~K\"uhn and T.~Rogers.
\newblock {Heterogeneous Micro-Structure of Percolation in Sparse Networks}.
\newblock {\em Europhys. Lett.}, 118:68003, 2017.

\bibitem{Lesk14}
J.~Leskovec and A.~Krevl.
\newblock {SNAP Datasets: Stanford Large Network Dataset Collection}.
\newblock {\tt http://snap.stanford.edu/data}, 2014.

\bibitem{AlbBarab02}
R.~Albert and A.-L. Barab\'asi.
\newblock {Statistical Mechanics of Complex Networks}.
\newblock {\em Rev.~Mod.~Phys.}, 74:47--97, 2002.

\bibitem{DorogMend03}
S.~N. Dorogovtsev and J.~F.~F Mendes.
\newblock {\em {Evolution of Networks: from Biological Networks to the Internet
  and WWW}}.
\newblock Oxford University Press, Oxford, 2003.

\bibitem{NewBk10}
M.~E.~J. Newman.
\newblock {\em {Networks: an Introduction, 2$^{\rm nd}$ Ed.}}
\newblock Oxford Univ. Press, Oxford, 2018.

\bibitem{Estrada2011}
E.~Estrada.
\newblock {\em {The Structure of Complex Networks: Theory and Applications}}.
\newblock Oxford University Press, Oxford, 2011.

\bibitem{Latora+2017}
V.~Latora, V.~Nicosia, and G.~Russo.
\newblock {\em {Complex Networks: Principles, Methods and Applications}}.
\newblock Cambridge University Press, Cambridge, 2017.

\bibitem{torrisi2021overcoming}
G.~Torrisi, A.~Annibale, and R.~K{\"u}hn.
\newblock Overcoming the complexity barrier of the dynamic message-passing
  method in networks with fat-tailed degree distributions.
\newblock {\em Phys. Rev. E}, 104(4):045313, 2021.

\bibitem{fink2021boolean}
T.~Fink and R.~Hannam.
\newblock Boolean composition restricts biological logics.
\newblock {\em arXiv preprint arXiv:2109.12551}, 2021.

\bibitem{gates2021effective}
A.~J Gates, R.~B Correia, X.~Wang, and L.~M Rocha.
\newblock The effective graph reveals redundancy, canalization, and control
  pathways in biochemical regulation and signaling.
\newblock {\em Proc. Natl. Acad. Sci.}, 118(12), 2021.

\bibitem{Torrisi+20}
G.~Torrisi, R.~K\"uhn, and A.~Annibale.
\newblock {Percolation in the Gene Regulatory Network}.
\newblock {\em JSTAT}, 083501:31p, 2020.

\bibitem{Hannam+19}
R.~Hannam, R.~K\"uhn, and A.~Annibale.
\newblock {Percolation in Bipartite Boolean Networks and its Role in Sustaining
  Life}.
\newblock {\em J. Phys. A}, 52:334002, 2019.

\bibitem{hertz1986memory}
JA~Hertz, G~Grinstein, and SA~Solla.
\newblock Memory networks with asymmetric bonds.
\newblock In {\em AIP Conference Proceedings}, volume 151, pages 212--218.
  American Institute of Physics, 1986.

\bibitem{gutfreund1988nature}
H~Gutfreund, JD~Reger, and AP~Young.
\newblock The nature of attractors in an asymmetric spin glass with
  deterministic dynamics.
\newblock {\em Journal of Physics A: Mathematical and General}, 21(12):2775,
  1988.

\bibitem{han2018trrust}
H.~Han, J.~Cho, S.~Lee, A.~Yun, H.~Kim, et~al.
\newblock Trrust v2: an expanded reference database of human and mouse
  transcriptional regulatory interactions.
\newblock {\em Nucleic Acids Res.}, 46(D1):D380--D386, 2018.

\bibitem{del2015dynamic}
G.~Del~Ferraro and E.~Aurell.
\newblock Dynamic message-passing approach for kinetic spin models with
  reversible dynamics.
\newblock {\em Phys. Rev. E}, 92(1):010102, 2015.

\bibitem{mezard2011exact}
M.~M{\'e}zard and J.~Sakellariou.
\newblock Exact mean-field inference in asymmetric kinetic ising systems.
\newblock {\em J. Stat. Mech.}, 2011(07):L07001, 2011.

\bibitem{gleeson2011high}
J.~P Gleeson.
\newblock High-accuracy approximation of binary-state dynamics on networks.
\newblock {\em Phys Rev. Let.}, 107(6):068701, 2011.

\bibitem{deplancke2012gene}
B.~Deplancke and N.~Gheldof.
\newblock {\em Gene regulatory networks: methods and protocols}.
\newblock Springer, 2012.

\bibitem{bitbol2016inferring}
A.~F. Bitbol, R.~S Dwyer, L.~J Colwell, and N.~Wingreen.
\newblock Inferring interaction partners from protein sequences.
\newblock {\em Proc. Natl. Acad. Sci. U.S.A.}, 113(43):12180--12185, 2016.

\bibitem{morcos2011direct}
F.~Morcos, A.~Pagnani, B.~Lunt, A.~Bertolino, D.~S Marks, C.~Sander,
  R.~Zecchina, J.~N Onuchic, T.~Hwa, and M.~Weigt.
\newblock Direct-coupling analysis of residue coevolution captures native
  contacts across many protein families.
\newblock {\em Proc. Natl. Acad. Sci. U.S.A.}, 108(49):E1293--E1301, 2011.

\bibitem{cavagna2014dynamical}
A.~Cavagna, I.~Giardina, F.~Ginelli, T.~Mora, D.~Piovani, R.~Tavarone, and A.~M
  Walczak.
\newblock Dynamical maximum entropy approach to flocking.
\newblock {\em Phys. Rev.E}, 89(4):042707, 2014.

\bibitem{cantwell2019message}
George~T Cantwell and Mark~EJ Newman.
\newblock Message passing on networks with loops.
\newblock {\em Proceedings of the National Academy of Sciences},
  116(47):23398--23403, 2019.

\bibitem{kirkley2021belief}
A.~Kirkley, G.~T Cantwell, and MEJ Newman.
\newblock Belief propagation for networks with loops.
\newblock {\em Science Advances}, 7(17):eabf1211, 2021.

\bibitem{lambiotte2019networks}
R.~Lambiotte, M.~Rosvall, and I.~Scholtes.
\newblock From networks to optimal higher-order models of complex systems.
\newblock {\em Nat. Phys.}, 15(4):313--320, 2019.

\bibitem{battiston2020networks}
F.~Battiston, G.~Cencetti, I.~Iacopini, V.~Latora, M.~Lucas, A.~Patania, J.-G.
  Young, and G.~Petri.
\newblock Networks beyond pairwise interactions: Structure and dynamics.
\newblock {\em Physics Reports}, 2020.

\bibitem{ghoshal2009random}
G.~Ghoshal, V.~Zlati{\'c}, G.~Caldarelli, and M.~EJ Newman.
\newblock Random hypergraphs and their applications.
\newblock {\em Physical Review E}, 79(6):066118, 2009.

\bibitem{petri2018simplicial}
G.~Petri and A.~Barrat.
\newblock Simplicial activity driven model.
\newblock {\em Physical review letters}, 121(22):228301, 2018.

\bibitem{KirkThir87a}
T.~R. Kirkpatrick and D.~Thirumalai.
\newblock {Dynamics of the Structural Glass Transition and the $p$-Spin
  Interaction Spin-Glass Model}.
\newblock {\em Phys. Rev. Lett.}, 58:2091--2094, 1987.

\bibitem{thomas1981relation}
Thomas R.
\newblock On the relation between the logical structure of systems and their
  ability to generate multiple steady states or sustained oscillations.
\newblock In {\em Numerical methods in the study of critical phenomena}, pages
  180--193. Springer, 1981.

\bibitem{plahte1995feedback}
E.~Plahte, T.~Mestl, and S.~W Omholt.
\newblock Feedback loops, stability and multistationarity in dynamical systems.
\newblock {\em J. Biol. Syst}, 3(02):409--413, 1995.

\bibitem{freeman2000feedback}
M.~Freeman.
\newblock Feedback control of intercellular signalling in development.
\newblock {\em Nature}, 408(6810):313--319, 2000.

\bibitem{zhang2012noise}
H.~Zhang, Y.~Chen, and Y.~Chen.
\newblock Noise propagation in gene regulation networks involving interlinked
  positive and negative feedback loops.
\newblock {\em PloS one}, 7(12):e51840, 2012.

\bibitem{chakravarty2021systematic}
S.~Chakravarty and A.~Csik{\'a}sz-Nagy.
\newblock Systematic analysis of noise reduction properties of coupled and
  isolated feed-forward loops.
\newblock {\em PLoS Comput. Biol.}, 17(12):e1009622, 2021.

\bibitem{mambuca2020dynamical}
A.~M. Mambuca, C.~Cammarota, and I.~Neri.
\newblock Dynamical systems on large networks with predator-prey interactions
  are stable and exhibit oscillations.
\newblock {\em Phys. Rev. E}, 105:014305, Jan 2022.

\bibitem{aurell2011message}
E.~Aurell and H.~Mahmoudi.
\newblock A message-passing scheme for non-equilibrium stationary states.
\newblock {\em J. Stat. Mech.}, 2011(04):P04014, 2011.

\bibitem{cardelli2016noise}
L.~Cardelli, A.~Csik{\'a}sz-Nagy, N.~Dalchau, M.~Tribastone, and
  M.~Tschaikowski.
\newblock Noise reduction in complex biological switches.
\newblock {\em Sci. Rep.}, 6(1):1--12, 2016.

\bibitem{campajola2021equivalence}
C.~Campajola, F.~Lillo, P.~Mazzarisi, and D.~Tantari.
\newblock On the equivalence between the kinetic ising model and discrete
  autoregressive processes.
\newblock {\em Journal of Statistical Mechanics: Theory and Experiment},
  2021(3):033412, 2021.

\bibitem{Anderson88}
J.~A. Anderson and E.~Rosenfeld, editors.
\newblock {\em {Neurocomputing: Foundations of Research}}.
\newblock MIT Press, Cambridge, 1988.

\end{thebibliography}

\appendix
\section{High noise regime}
\label{sec:high noise}
In the high noise ($\beta\rightarrow 0$ limit), the expansion of $\Phi_{\beta}(x)$ in \Eref{eq: result cavity} to first order in $\beta x$ \footnote{Note that $\Phi_{\beta}(x)= \Phi_1(\beta x)$, both for thermal noise, and for Gaussian noise, if we equate $\beta=\sigma$ in the Gaussian case} gives
\begin{equation}
P_i(t+1) \approx \frac{1}{2}+\beta \Phi_1'(0) \left( \sum_j J_{ij} P_j(t)  -\vartheta_i \right)\,.
\end{equation}
Using $P_j(t) =\frac{1}{2}+\mathcal{O}(\beta)$,  we obtain that 
\begin{equation}
P_i(t+1) =\frac{1}{2}+ \beta  \Phi_1'(0)\left(\frac{1}{2}\sum_j^N J_{ij}-\vartheta_i\right)+\mathcal{O}(\beta^2)\,,
\label{eq: approx cavity high T}
\end{equation}
which is independent of the time-step $t$.
The result of \Eref{eq: approx cavity high T} states that in the large noise limit, the  node activation of node $i$, $P_i$, is solely determined by the imbalance of the interactions with $i$'s predecessors, i.e. by $\sum_j J_{ij}$.  The mean node activation probability over sites $\langle P\rangle = \sum_i P_i/N$ reduces to 
\begin{equation}
\langle P\rangle = \frac{1}{2} + \beta  \Phi_1'(0)\left(\frac{1}{2N}\sum_{i,j} J_{ij}-\frac{1}{N}\sum_i\vartheta_i\right)+\mathcal{O}(\beta^2) \,,
\label{eq: mean_cavity_high_T}
\end{equation}
where the $\mathcal{O}(\beta)$ correction depends only on the macroscopic quantities $\sum_{i,j}J_{ij}/N$ and $\sum_i\vartheta_i/N$. This result is compatible with the expansion around the paramagnetic phase done in ref. \cite{neri2009cavity}.

Similarly, the high noise approximation for the AND gate dynamics discussed  in \Eref{eq:non_linear_prob2} gives
\begin{equation}
P_i \approx \frac{1}{2}+\beta \Phi_1'(0)\left[\sum_\mu J_{i\mu}\mathcal{P}_\mu-\vartheta_i\right]\,,
\label{eq:high_T_AND}
\end{equation}where 
\begin{equation}
\mathcal{P}_\mu = \left(\frac{1}{2}\right)^{c_\mu}+\beta \Phi_1'(0)\sum_{j\in\partial_\mu} \sum_\nu J_{j\nu}\left(\frac{1}{2}\right)^{c_\nu}+\mathcal{O}(\beta^2) \,.
\end{equation}
In this case, the average node activation probability is therefore given by 
\begin{equation}
\langle P\rangle = \frac{1}{2} + \beta  \Phi_1'(0)\left[\frac{1}{N}\sum_{i,\mu} J_{i\mu}\left(\frac{1}{2}\right)^{c_\mu}-\frac{1}{N}\sum_i\vartheta_i\right]+\mathcal{O}(\beta^2)\ .
\label{eq: mean_cavity_high_T_AND}
\end{equation}
In the high noise limit,  the average activation  probability of factor nodes  converges to
\begin{equation}
\langle \mathcal{P}\rangle = \sum_{c_\mu} P(c_\mu) \left(\frac{1}{2}\right)^{c_\mu} \defeq G^{\mathrm{C}}_0(1/2)\ ,\quad \mbox{as }\beta\to 0. 
\end{equation}
with $G^{\mathrm{C}}_0(x)$ the generating function of the in-degree of factor nodes. This demonstrates, among other things, that the signalling through AND gates  with large in-degrees is effectively  suppressed  in the high noise limit. Thus, factor nodes with large in-degree  may operate as noise filters. 
\section{Derivation of the One-Time Approximation}
\label{sec:derivation OTA}
\paragraph{Dynamic cavity trajectory with bi-directional links}
For completeness, we include in this appendix 
a brief derivation of the OTA approximation from the dynamic cavity equations, mainly following Ref.\,\cite{neri2009cavity,aurell2012dynamic}. We also compare different closure schemes of the OTA which have been proposed in the literature and propose a new version as an alternative.

Let us define $P_i\left(\bm{n}_i^{0,\dots,t}|\vartheta_i\right)$ the probability of the trajectory  $\bm{n}_i^{0,\dots,t}\defeq (n_i(0) \dots n_i(t))$  of node $i$ over the time span $0,\dots,t$, 
and $P_j^{(i)}\left(\bm{n}_j^{0,\dots,t-1}|\bm{\vartheta}_j^{(i)0,\dots,t-2}\right)$ the conditional probability of the trajectory $\bm{n}_j^{0,\dots,t-1}\defeq (n_j(0) \dots n_j(t-1))$ of node $j$, given the history of node $i$,
in the cavity graph where node $i$ and its links have been removed, so that the past values of node $i$ act as external thresholds. 
Their contribution can be combined with those of the constant threshold $\vartheta_j$, to define  time-dependent thresholds 
$\bm{\vartheta}_j^{(i)0,\dots,t-1}\defeq
(\vartheta_j^{(i)}(0) \dots \vartheta_j^{(i)}(t-1))$ with  \begin{equation}
\vartheta_j^{(i)}(s)= \vartheta_j-J_{ji}n_i(s)\,,
\label{eq:retarded_local_cavity_field}
\end{equation}

where $J_{ji}n_i^{s}$ is the signal received by $j$ from $i$ as a retarded interaction.  
Above, we have used the notation with time superscript to indicate the collection of variables over the trajectory. In what follows, 
to ease the notation, we will also use superscripts for single time quantities, e.g. $n_i(s)=n_i^{s}$ and $\vartheta_j^{(i)}(s)=\vartheta_j^{(i),s}$.

The probability of a  single-node trajectory  does not factorise in the time-steps, but one needs to solve the set of equations, see Ref.\,\cite{neri2009cavity,aurell2011message}
\begin{multline}
P_i\left(\bm{n}_i^{0,\dots,t}|\vartheta_i
\right)=\sum_{\bm{n}^0_{\partial_i}}\dots \sum_{\bm{n}^{t-1}_{\partial_i}}\left[\prod_{s=0}^{t-1}  W\left[ n_i^{s+1}|h_i(\bm{n}_{\partial_i}^s),\vartheta_i
\right]\right]\\
~~~~~~~~~~~~~~~~\times \prod_{j\in \partial_i} P_j^{(i)}\left(\bm{n}_j^{0,\dots,t-1}|\bm{\vartheta}_j
-  J_{ji}\bm{n}_i^{0,\dots,t-2}\right)P_i(n_i^0)\,,
\label{eq:cavity_bidirectional}
\end{multline}
\begin{multline}
P_j^{(i)}\left(\bm{n}_j^{0,\dots,t-1}|\bm{\vartheta}
^{(i)0,\dots,t-2}_j
\right)= \sum_{\bm{n}^0_{\partial_j\setminus i}}\dots \sum_{\bm{n}^{t-2}_{\partial_j\setminus i}}\left[ \prod_{s=0}^{t-2} W\left[ n_j^{s+1}|h_j^{(i)}(\bm{n}_{\partial_j}^s),\vartheta_j^{(i),s}\right]\right]
\\
~~~~~~~~~~~~~~~~\times\prod_{\ell\in \partial_j\setminus i} P_\ell^{(j)}\left(\bm{n}_\ell^{0,\dots,t-2}|\bm{\vartheta}_\ell
-  J_{\ell j}\bm{n}_j^{0,\dots,t-3}\right)P_j(n_j^0)\,,
\label{eq:cavity_bidirectional2}
\end{multline}
where $\bm{\vartheta}_j=(\vartheta_j,\dots, \vartheta_j)$ 
is a constant vector with identical 
entries $\vartheta_j$.
Note that for any node $j\in \partial_i$, i.e. such that $J_{ij}\neq 0$, one has $J_{ji}\neq 0$ only if the link $(i,j)$ is bidirectional, so the
retarded self-interaction terms in \Eref{eq:retarded_local_cavity_field} are absent for 
unidirectional links. 
The transition probability $W\left[ n_j^{s+1}|h_j^{(i)}(\bm{n}_{\partial_j}^s),\vartheta_j^{(i),s}\right] $ indicates the conditional probability of the state $n_j^{s+1}$ at time $s+1$ given the values of the  cavity  local field $h_j^{(i)}(\bm{n}_{\partial_j}^s)$, 
\begin{equation}
 h_j^{(i)}(\bm{n}_{\partial_j}^s)\defeq \sum_{\ell\in \partial_j \setminus { i}} J_{j\ell}n_\ell^s\,,
\end{equation}
and the cavity threshold $\vartheta_j^{(i),s}$.
The detailed function form  of $W\left[ n_j^{s+1}|h_j^{(i)}(\bm{n}_{\partial_j}^s),\vartheta_j^{(i),s}\right] $ depends on the choice of the noise distribution. In analogy with the above we adopt a thermal noise model for which
\begin{equation}
W\left[ n_j^{s+1}|h_j^{(i)}(\bm{n}_{\partial_j}^s),\vartheta_j^{(i),s}\right] =  \frac{1}{2}\left\lbrace 1+(2n_j^{s+1}-1)\tanh \left[\frac{\beta}{2}\left(h_j^{(i)}(\bm{n}_{\partial_j}^s)-\vartheta_j^{(i),s}\right) \right]\right\rbrace\,.
\end{equation}
 In \Eref{eq:cavity_bidirectional} and \Eref{eq:cavity_bidirectional2}  the probability of the state of node $i$  depends on the states of neighbours $\partial_i$, which in turn depend parametrically on the node $i$ through the cavity thresholds. This feedback effect  is responsible for  the  non-Markovian structure of \Eref{eq:cavity_bidirectional} and \Eref{eq:cavity_bidirectional2}, entailing that  the computational complexity for a trajectory of time length $t$ grows as $2^{t}$, making the evaluation feasible  only for relatively few time-steps. 
 
 From now on, we focus on  the single site cavity probability and we discuss the quantity $P_j^{(i)}\left(\bm{n}_j^{0,\dots,t}|\bm{\vartheta}^{(i),0,\dots,t-1}_j\right)$ for the trajectory up to time $t$ and not up to time $t-1$ as done in \Eref{eq:cavity_bidirectional2} to simplify the notation of the lagged time dependence.  
 A pragmatic approach  is to  assume that  the single site  cavity probability  trajectory $P_j^{(i)}\left(\bm{n}_j^{0,\dots,t}|\bm{\vartheta}^{(i),0,\dots,t-1}_j\right)$  factorises in a Markovian fashion under the  one-time approximation (OTA)\cite{neri2009cavity} 
\begin{equation}
P_j^{(i)}\left(\bm{n}_j^{0,\dots,t}|\bm{\vartheta}^{(i), 0,\dots,t-1}_j\right)  \approx P^{(i)}_j(n_j^0)\prod_{s=0}^{t-1} P^{(i)}_j(n_j^{s+1}|\vartheta^{(i),s}_j)\,.
\label{eq:factorisation_trajectory}
\end{equation}
Substituting \Eref{eq:factorisation_trajectory} into  \Eref{eq:cavity_bidirectional2}  
\begin{multline}
P_j^{(i)}\left(\bm{n}_j^{0,\dots,t}|\bm{\vartheta}^{(i), 0,\dots,t-1}_j\right) = \left[ \sum_{\bm{n}^{t-1}_{\partial_j\setminus i}}W\left[ n_j^t|h_j^{(i)}(\bm{n}_{\partial_j}^{t-1}),\vartheta_j^{(i), t-1}\right]\prod_{\ell\in \partial_j\setminus i} P_\ell^{(j)}\left(n_\ell^{t-1}|\vartheta_\ell-J_{\ell j}n_j^{ t-2}\right)\right] \\
~~~~~~~~~~~~~~~~\times P_j^{(i)}\left(\bm{n}_j^{0,\dots,t-1}|\bm{\vartheta}
^{(i)0,\dots,t-2}_j
\right)\,,
\label{eq:cavity_OTA_joint_right}
\end{multline}
where we have grouped together the terms of \Eref{eq:factorisation_trajectory} at the previous time-steps
\begin{multline}
P_j^{(i)}\left(\bm{n}_j^{0,\dots,t-1}|\bm{\vartheta}
^{(i)0,\dots,t-2}_j
\right) = 
\sum_{\bm{n}^0_{\partial_j\setminus i}}\dots \sum_{\bm{n}^{t-2}_{\partial_j\setminus i}}\left[ \prod_{s=0}^{t-2} W\left[ n_j^{s+1}|h_j^{(i)}(\bm{n}_{\partial_j}^s),\vartheta_j^{(i),s}\right]\right]
\\
~~~~~~~~~~~~~~~~\times\prod_{\ell\in \partial_j\setminus i}\left[\prod_{s=0}^{t-3} P_\ell^{(j)}\left(n_\ell^{s+1}|\vartheta_\ell-  J_{\ell j}n_j^{s}\right)P_\ell(n_\ell^0)\right]P_j(n_j^0)\,.
\label{eq:cavity_OTA_factor_right}
\end{multline}

\begin{figure}
\centering
\includegraphics[width = 0.49\textwidth]{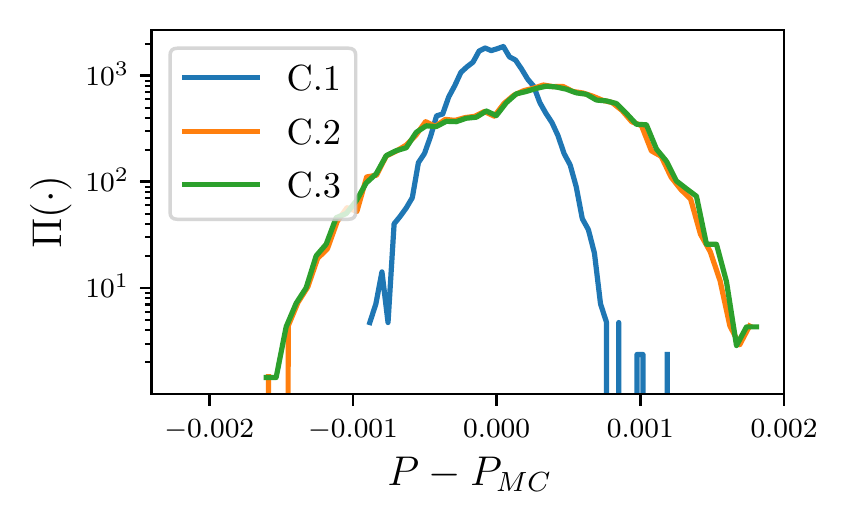}
\includegraphics[width = 0.49\textwidth]{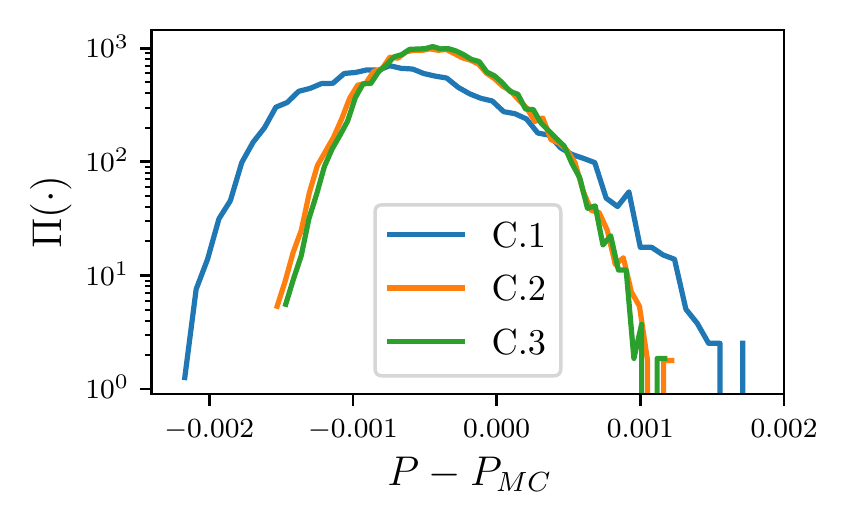}
\caption{Comparison of different closures in the OTA. C.1 is the closure of Ref.\,\cite{aurell2012dynamic}. C.2 is the closure of Ref.\,\cite{zhang2012inference}, C.3 is the closure of \Eref{eq:closure_averaged}. The distribution of the difference between the probability obtained using OTA and the probability obtained using Monte Carlo simulations. Network with symmetric (left) and antisymmetric interactions (right). The network  belongs to the  random regular graph class with $k^{\mathrm{in}}=k^{\mathrm{out}}=3$. Simulations are obtained through an average of over 500 trajectories. Parameters are $N = 10000$, $T/J = 1$, and  $\vartheta = 0$, $t_s = 10^4$ steps.}
\label{fig:comparing_OTA}
\end{figure}

Applying the OTA factorization to the rightmost  factor in  \Eref{eq:cavity_OTA_joint_right}, i.e. \Eref{eq:cavity_OTA_factor_right}, one obtains 
\begin{multline}
P_j^{(i)}\left(\bm{n}_j^{0,\dots,t}|\bm{\vartheta}^{(i), 0,\dots,t-1}_j\right) = \left[ \sum_{\bm{n}^{t-1}_{\partial_j\setminus i}}W\left[ n_j^t|h_j^{(i)}(\bm{n}_{\partial_j}^{t-1}),\vartheta_j^{(i), t-1}\right]\prod_{\ell\in \partial_j\setminus i} P_\ell^{(j)}\left(n_\ell^{t-1}|\vartheta_\ell-J_{\ell j}n_j^{ t-2}\right)\right]\\ ~~~~~~~~~~~~~~~~~~\times\left[ \prod_{s=0}^{t-2}  P^{(i)}_j(n_j^{s+1}|\vartheta^{(i),s}_j) \right]P_j(n_j^0)\,.
\label{eq:OTA_approx}
\end{multline}
This expression is particularly suitable to compute the marginal over the trajectory of node $j$, i.e.,
\begin{equation}
P_j^{(i)}\left( n_j^t|\bm{\vartheta}_j^{(i),0,\dots,t-1}\right)= \sum_{\bm{n}_j^{0\dots t-1}}P_j^{(i)}\left(\bm{n}_j^{0,\dots,t}|\bm{\vartheta}^{(i), 0,\dots,t-1}_j\right)\,.
\label{eq:marginal_cav}
\end{equation}
Thanks to the factorization of the last term in \Eref{eq:OTA_approx}, the sum over $\bm{n}_j^{0\dots t-3,t-1}$ is trivial, as the term in the first square brackets does not depend on these variables, and conditional probabilities are normalised to one.
\begin{multline}
P_j^{(i)}\left( n_j^t|\vartheta_j^{(i), t-1},\vartheta_j^{(i), t-3}\right)= \\ \sum_{n_j^{t-2}} \left[ \sum_{\bm{n}^{t-1}_{\partial_j\setminus i}}W\left[ n_j^t|h_j^{(i)}(\bm{n}_{\partial_j}^{t-1}),\vartheta_j^{(i),t-1}\right]\prod_{\ell\in \partial_j\setminus i} P_\ell^{(j)}\left(n_\ell^{t-1}|\vartheta_\ell- J_{\ell j}n_j^{ t-2}\right)\right] P^{(i)}_j(n_j^{t-2}|\vartheta^{(i),t-3}_j)
\label{eq:marginal_cavity_2_times}
\end{multline}
In  \Eref{eq:marginal_cavity_2_times} the probability of node $j$ depends on the two cavity thresholds $\vartheta_j^{(i), t-1},\vartheta_j^{(i), t-3}$. This expression is directly derived from the OTA factorisation \Eref{eq:factorisation_trajectory}. The one-step object $P_j^{(i)}\left( n_j^t|\vartheta_j^{(i), t-1}\right)$ can be obtained imposing  a ``closure'' condition that enforces the Markovian behaviour, and different choices could be made. In \cite{neri2009cavity,aurell2011message} the authors assume that, in the stationary state, the  cavity threshold  does not have an explicit time dependence $\vartheta_j^{(i),s}\approx \vartheta_j^{(i)}$ $\forall s$. In \cite{zhang2012inference} instead, the cavity term is approximated by the non-cavity marginal $P^{(i)}_j(n_j^{t-2}|\vartheta^{(i),t-3}_j)\approx P_j(n_j^{t-2}|\vartheta_j)$. Another approach is to truncate the retarded dependence by taking an average over the state $n_i^{t-3}$, which we detail below. Remembering from \Eref{eq:retarded_local_cavity_field} that the $\vartheta_j^{(i),s}=\vartheta_j-J_{ji}n_i^{s}$, then
\begin{equation}
P(\vartheta_j^{(i), t-3}) =\sum_{n_i^{t-3}}\delta_{\vartheta_j^{(i), t-3},\vartheta_j-J_{ji}n_i^{t-3}}P(n_i^{t-3}|\vartheta_i)\,,
\end{equation}   with $\delta$ indicating the Kronecker delta. The one-time cavity marginal appearing in \Eref{eq:factorisation_trajectory} is given by  
\begin{equation}
P_j^{(i)}\left( n_j^t|\vartheta_j^{(i), t-1}\right) =\sum_{n_i^{t-3}}P_j^{(i)}\left( n_j^t|\vartheta_j^{(i), t-1},\vartheta_j-J_{ji}n_i^{t-3}\right)P_i(n_i^{ t-3}|\vartheta_i)\,.
\label{eq:cavity_marginalisation}
\end{equation}
Substituting \Eref{eq:marginal_cavity_2_times} in \Eref{eq:cavity_marginalisation}, the expression for the one-step probability becomes
\begin{equation}
P_j^{(i)}\left( n_j^t|\vartheta_j^{(i), t-1}\right)=  \sum_{n_j^{t-2}} \left[ \sum_{\bm{n}^{t-1}_{\partial_j\setminus i}}W\left[ n_j^t|h_j^{(i)}(\bm{n}_{\partial_j}^{t-1}),\vartheta_j^{(i), t-1}\right]\prod_{\ell\in \partial_j\setminus i} P_\ell^{(j)}\left(n_\ell^{t-1}|\vartheta_\ell- J_{\ell j}n_j^{ t-2}\right)\right] P^{(i)}_j(n_j^{t-2}|\vartheta_j)
\label{eq:marginal_cavity_1_time}
\end{equation}
with
\begin{equation}
P^{(i)}_j(n_j^{t-2}|\vartheta_j) =\sum_{n_i^{t-3}} P^{(i)}_j(n_j^{t-2}|\vartheta_j-J_{ji}n_i^{t-3})P_{i}(n_i^{t-3}|\vartheta_i)
\label{eq:closure_averaged}
\end{equation}
We compare the different  OTA closure conditions in \Eref{eq:marginal_cavity_2_times}, which we denote:
\begin{itemize}
\item C.1 corresponding to the assumption $P^{(i)}_j(n_j^{t-2}|\vartheta^{(i),t-3}_j)\approx P^{(i)}_j(n_j^{t-2}|\vartheta^{(i),t-1}_j)$, from Ref.\,\cite{aurell2012dynamic},
\item C.2 corresponding to the assumption $P^{(i)}_j(n_j^{t-2}|\vartheta^{(i),t-3}_j)\approx P_j(n_j^{t-2}|\vartheta_j)  $, from Ref.\,\cite{zhang2012inference},
\item C.3 corresponding to the assumption $P^{(i)}_j(n_j^{t-2}|\vartheta^{(i),t-3}_j)\approx P^{(i)}_j(n_j^{t-2}|\vartheta_j)$ using the definition in  \Eref{eq:closure_averaged}.
\end{itemize}
We evaluate the distribution of the stationary single-node activation probability  for the  OTA expressions corresponding to closures C.1, C.2, and C.3. 
We benchmark the theory with Monte Carlo simulations for  symmetric and antisymmetric networks, and we show in Fig.\,\ref{fig:comparing_OTA} the distribution of the distance between theory and simulation for the single-site activation probabilities. Our results indicate that in the symmetric case closure C.1 outperforms both C.2 and C.3, which is  expected since the expression associated with C.1 admits the equilibrium solution of belief propagation, as discussed in Ref.\,\cite{neri2009cavity}. In the antisymmetric case instead,  the relative performances of the methods are changed,  with closure C.1 providing the worst result roughly by a factor of two compared to closures C.2 and C.3. We use the mean square distance between OTA closures and simulation to quantify the error in Tab.\,\ref{tab:comparison_OTA}. Our results indicate that closures C.2  and C.3 are less dependent on the network symmetry, while C.1 presents important variations between the two cases examined.    From a computational point of view, closures C.1  and C.2 are comparable and they are simpler to perform compared to C.3, since in closure C.3,  the Eq.\,\eqref{eq:closure_averaged} needs to be performed for every link and for every time-step of interest.  Hence we adopt  closure C.2  since it is simpler to implement than C.3, and our results suggest it gives the same performances as  closure C.3. 

\begin{table}[]
\centering
\caption{Error, as defined in \Eref{eq: error generic}, of different closure schemes for symmetric, antisymmetric and uncorrelated interactions. Same parameters as in Fig.\,\ref{fig:comparing_OTA}.}
\begin{tabular}{|llll|}\hline
              & C.1      & C.2      & C.3      \\\hline
symmetric     & \num{2.3e-04} & \num{6.0E-04} & \num{6.2E-04} \\
antisymmetric & \num{8.6E-04} & \num{5.5E-04} & \num{5.3E-04}\\
uncorrelated  & \num{3.8E-04} & \num{3.4E-04} & \num{3.4E-04}   \\\hline
\end{tabular}

\label{tab:comparison_OTA}
\end{table}

\end{document}